\pdfoutput=1
\documentclass[aps,twocolumn, floatfix, superscriptaddress, prx]{revtex4}
\usepackage{graphicx}
\DeclareGraphicsExtensions{.pdf}
\usepackage{amsmath,amssymb,bbold,bm,color}
\usepackage{float}
\usepackage{epstopdf}
\usepackage{tabularx}
\usepackage{braket}
\usepackage{booktabs}
\usepackage{array}
\newcolumntype{L}[1]{>{\raggedright\let\newline\\\arraybackslash\hspace{0pt}}m{#1}}
\newcolumntype{C}[1]{>{\centering\let\newline\\\arraybackslash\hspace{0pt}}m{#1}}
\newcolumntype{R}[1]{>{\raggedleft\let\newline\\\arraybackslash\hspace{0pt}}m{#1}}
\usepackage{calrsfs}
\DeclareMathAlphabet{\pazocal}{OMS}{zplm}{m}{n}
\newcommand{\sgn}{\text{sgn}}

\begin{document}

\title{Magnon quantum anomalies in Weyl ferromagnets}

\author{Tianyu Liu}
\affiliation{Department of Physics and Astronomy, University of British Columbia, Vancouver, BC, Canada V6T 1Z1}
\affiliation{Quantum Matter Institute, University of British Columbia, Vancouver BC, Canada V6T 1Z4}

\author{Zheng Shi}
\affiliation{Dahlem Center for Complex Quantum Systems and Physics Department,
Freie Universit\"at Berlin, Arnimallee 14, 14195 Berlin, Germany}

\begin{abstract} 
When subjected to parallel electric field $\bm E$ and magnetic field $\bm B$, Weyl semimetals exhibit the exotic transport property known as the chiral anomaly due to the pumping of electrons between Weyl cones of opposite chiralities. When one or both electromagnetic (EM) fields are replaced by strain-induced chiral pseudo-electromagnetic (pseudo-EM) fields, other types of quantum anomalies occur. In the present paper, we will show that such quantum anomalies can be reproduced in a completely different system -- a Weyl ferromagnet whose magnonic structure remarkably encodes Weyl physics. By analytical and numerical calculations, we will show that the magnon bands can be Landau-quantized by either an inhomogeneous electric field $\bm E$ or a chiral pseudo-electric field $\bm e$ induced by a torsional strain, and magnons can be pumped along bands by either an inhomogeneous magnetic field $\bm B$ or a chiral pseudo-magnetic field $\bm b$ due to a dynamic uniaxial strain. We list the magnon quantum anomalies and the associated anomalous spin and heat currents in the Weyl ferromagnet, and show they are distinct from their Weyl semimetal counterparts.
\end{abstract}

\date{\today}
\maketitle

\section{Introduction}
A magnon is a bosonic collective excitation of electron spins on a crystal lattice, carrying spin, magnetic dipole moment, and heat \cite{bloch1930}. It can be quantum-mechanically characterized by a quantized spin wave \cite{holstein1940} and experimentally detected by neutron scattering \cite{brockhouse1957}. Though charge neutral, a magnon can be manipulated by electric fields and magnetic fields through the Aharonov-Casher effect \cite{aharonov1984} and the Zeeman effect, respectively. Due to their unique reaction to electromagnetic (EM) fields, magnons stand out from other bosons (photons, phonons, polarons, etc.) but are akin to electrons. Consequently, magnons are proposed to mimic some electronic transport properties such as the chiral anomaly \cite{su2017a, su2017b}, the magnon Josephson effect \cite{nakata2014}, the spin Hall effect \cite{zyuzin2018}, the Wiedemann-Franz law \cite{nakata2017, nakata2015}, the thermal Hall effect \cite{laurell2018, matsumoto2014, zhang2013, matsumoto2011, katsura2010, onose2010}, the spin Seebeck effect \cite{xiao2010, rezende2014, uchida2011, uchida2010, uchida2008}, the spin Peltier effect \cite{kovalev2012, flipse2014}, and the spin Nernst effect \cite{cheng2016, zyuzin2016, kovalev2016, shiomi2017}.

On the other hand, electronic transport does not necessarily require the participation of EM fields or temperature gradients. One example is that relativistic electrons can be manipulated by elastic strains. In graphene, for instance, the crystal lattice deformation due to strain shifts the 2D Dirac electrons in the momentum dimension \cite{guinea2010}. Thus it gives rise to a pseudo-magnetic field, generating Landau levels that can be observed in scanning tunneling microscopy (STM) \cite{levy2010}. In Dirac/Weyl semimetals, it has been theoretically proposed that a similar pseudo-magnetic field couples to 3D Dirac/Weyl particles \cite{liu2017a, sumiyoshi2016, pikulin2016, grushin2016, cortijo2015}, giving rise to quantum oscillations \cite{liu2017a} and the chiral torsional effect \cite{sumiyoshi2016, pikulin2016} without magnetic field. Remarkably, when a suitable dynamic strain is applied, a pseudo-electric field emerges, producing a chiral anomaly in Weyl semimetals in the complete absence of EM fields \cite{pikulin2016}.

Recent works on Dirac/Weyl superconductors showed that strain can manipulate \textit{charge neutral} relativistic Bogoliubov quasiparticles, resulting in the Wiedemann-Franz law, quantum oscillations, and anomalous spin and heat currents \cite{liu2017b, matsushita2018}. These discoveries shed new light on novel magnon transport with only electric (magnetic) fields or even in the complete absence of EM fields as long as a suitable strain is present. It has been reported that strain is able to Landau-quantize 2D Weyl magnons hosted by ``magnon graphene" and leads to a non-quantized Hall viscosity \cite{ferreiros2018}. Since 3D Weyl magnons have been predicted to exist in pyrochlore magnets \cite{su2017b, mook2016, li2016, jian2018}, double perovskite magnets \cite{li2017} and multilayer magnets \cite{su2017a, zyuzin2018, owerre2018}, it is natural to ask how strain engages in the transport of 3D Weyl magnons harbored by these Weyl magnets.

In this paper, we answer this question by a combination of analytical calculations and numerical simulations. We show that a static torsional strain can induce a pseudo-electric field to Landau-quantize the 3D Weyl magnons, and a dynamic uniaxial strain can induce a pseudo-magnetic field to pump the magnons. We also demonstrate that these elastic strain-induced pseudo-EM fields result in novel magnon transport in the form of quantum anomalies. To support these findings, we organize the paper as follows. In Sec.~\ref{s2}, we formulate the model of a multilayer Weyl ferromagnet and discuss its magnon band structure and topology. In Sec.~\ref{s3}, we study the magnon band structures of the Weyl ferromagnet under an electric field and under a static torsional strain-induced pseudo-electric field respectively. We also study the magnon dynamics due to either a magnetic field or a dynamic uniaxial strain-induced pseudo-magnetic field. In Sec.~\ref{s4}, we derive the magnon quantum anomalies and the associated anomalous spin and heat currents under different combinations of EM fields and pseudo-EM fields. In Sec.~\ref{s5}, we derive the field (gradient) dependence of the anomalous spin and heat currents in various magnon quantum anomalies, and establish a duality to the anomalous electric current in the chiral magnetic effect \cite{fukushima2008, liqiang2016} and the chiral torsional effect \cite{pikulin2016, sumiyoshi2016} of Weyl semimetals. In Sec.~\ref{s6}, we discuss the proposals for experimentally measuring magnon quantum anomalies by atomic force microscopy (AFM) force sensing. Section~\ref{s7} concludes the paper, proposes several promising materials for the implementation of magnon quantum anomalies, and envisages a few worthwhile directions based on the current work.

\section{Multilayer model of Weyl ferromagnet}
\label{s2}
\begin{figure}[htb]
\includegraphics[width = 8cm]{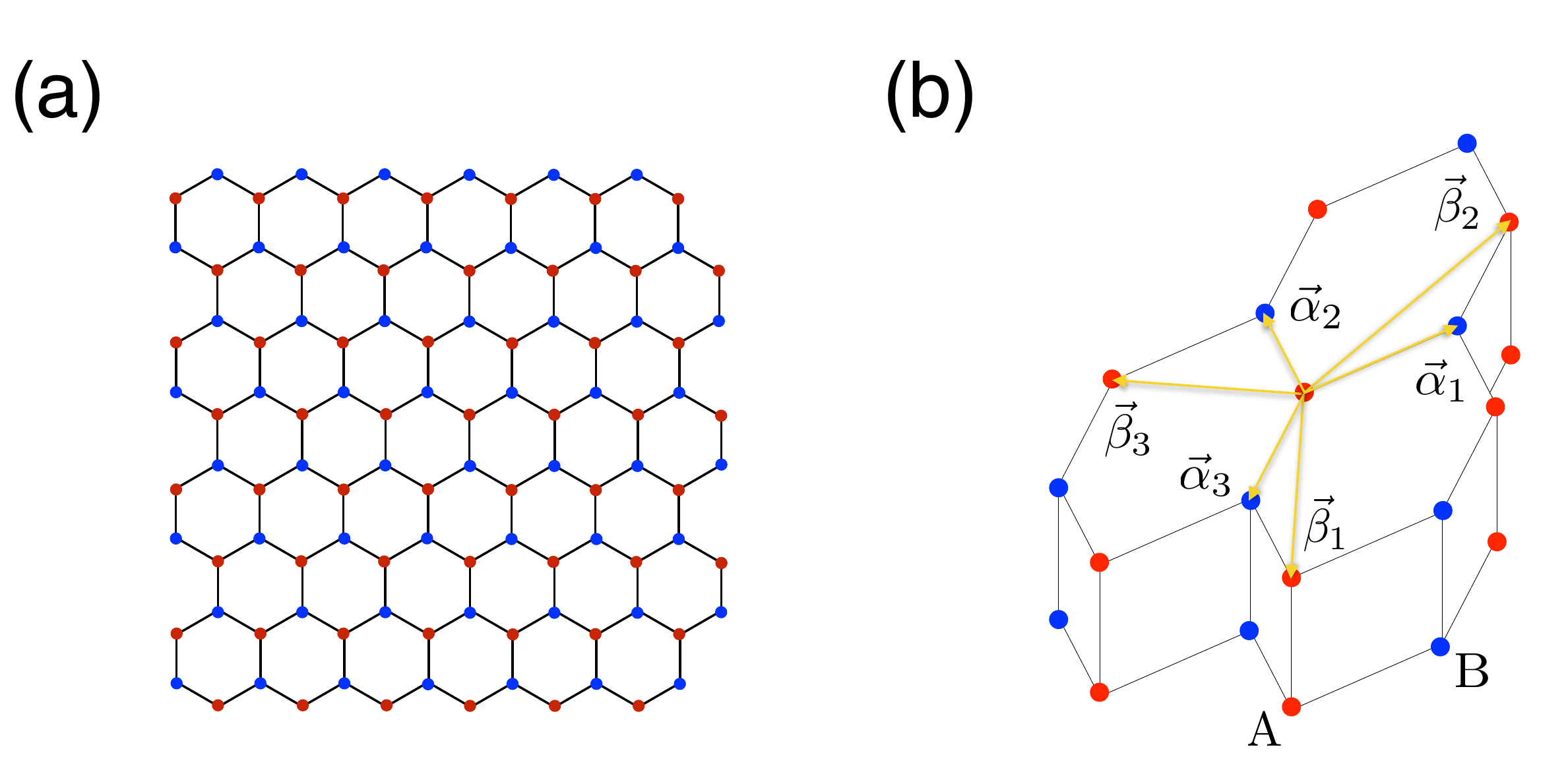}
\caption{Schematic plot for the Weyl ferromagnet multilayer. (a) 2D honeycomb ferromagnet sheet. The Weyl ferromagnet multilayer is constructed by stacking many sheets in the $z$ direction.  (b) Conventional crystal cells of the Weyl ferromagnet with in-plane nearest (second nearest) neighbors connected by $\bm \alpha_i$ ($\bm \beta_i$), $i=1,2,3$. } \label{fig_1}
\end{figure}

We consider a multilayer Weyl ferromagnet model proposed in Ref.~\cite{su2017a} but amend the model with an inter-layer next nearest neighbor interaction. It will be shown in Sec.~\ref{s3b} that such an interaction is necessary for the strain engineering for Landau quantization. The building block of this model is a honeycomb ferromagnet layer of spins of size $S$ (Fig.~\ref{fig_1}(a)), which has been experimentally realized in CrI$_3$ \cite{huang2017} and proposed in other compounds \cite{pershoguba2018}. Multiple layers are then stacked in the $z$ direction (Fig.~\ref{fig_1}(b)). The in-plane nearest neighbors are labelled by $\bm \alpha_1= \frac{\sqrt{3}}{2} a \hat x + \frac{1}{2} a \hat y$, $\bm \alpha_2= -\frac{\sqrt{3}}{2} a \hat x + \frac{1}{2} a \hat y$, and $\bm \alpha_3 = -a\hat y$, where $a$ is the lattice constant of the honeycomb lattice; the in-plane next nearest neighbors are labelled by $\bm \beta_1 =\bm \alpha_3 -\bm \alpha_2 $, $\bm \beta_2 = \bm \alpha_1 -\bm \alpha_3$, and $\bm \beta_3 =\bm \alpha_2 -\bm \alpha_1$. The inter-layer spacing (out-of-plane direction lattice constant) is denoted as $\delta_z$. We choose the unit cells of the honeycomb layer such that each contains an $A$ site and a $B$ site connected by vector $\bm \alpha_1$. The Heisenberg Hamiltonian of the Weyl ferromagnet reads
\begin{equation} \label{H_Heisenberg}
H= H_0 + H_1 + H_2 + H_z + H_D+Z,
\end{equation}
with its components listed explicitly below
\begin{equation*}
H_0 = -\sum_{\bm R_\perp, z} \sum_{\mu=A,B} K_\mu S^z_\mu(\bm R_\perp, z) S^z_\mu(\bm R_\perp, z),
\end{equation*}
\begin{equation*}
H_1 = -\sum_{\bm R_\perp, z} \sum_{i} J_1(\bm \alpha_i) \bm S_A(\bm R_\perp, z) \cdot \bm S_B(\bm R_\perp + \bm \alpha_i - \bm \alpha_1, z),
\end{equation*}
\begin{multline*}
H_2 =- \sum_{\bm R_\perp, z} \sum_{i} \sum_{s=\pm 1} J_2 (\bm \alpha_i + s \delta_z \hat z) \times \\ \bm S_A(\bm R_\perp, z) \cdot \bm S_B(\bm R_\perp + \bm \alpha_i - \bm \alpha_1, z + s\delta_z)
\end{multline*}
\begin{equation*}
H_z = - \sum_{\bm R_\perp, z}  \sum_{\mu=A,B} J_\mu( \delta_z \hat z) \bm S_\mu(\bm R_\perp, z) \cdot \bm S_\mu(\bm R_\perp, z + \delta_z),
\end{equation*}
\begin{equation*}
H_D = \sum_{\bm R_\perp, z} \sum_{i} \sum_{\mu=A,B} \bm D_\mu \cdot [\bm S_\mu (\bm R_\perp,z) \times \bm S_\mu (\bm R_\perp + \bm \beta_i,z)],
\end{equation*}
\begin{equation*}
Z=-g\mu_BB_z\sum_{\bm R_\perp,z} \sum_{\mu=A, B} S_\mu^z(\bm R_\perp, z),
\end{equation*}
where $\bm R_\perp$ labels the unit cells of a honeycomb layer, and $z$ is the layer index. $H_0$ is an on-site interaction term with strength $K_{A(B)}$ on $A(B)$ sites; $H_1$ is the intra-layer nearest neighbor interaction with strength $J_1(\bm \alpha_i)$ between $A$ and $B$ sites connected by vector $\bm \alpha_i$; $H_2$ is the inter-layer next nearest neighbor interaction with strength $J_2(\bm \alpha_i \pm \bm \delta_z)$ between $A$ and $B$ sites connected by vector $\bm \alpha_i \pm \bm \delta_z$; $H_z$ is the inter-layer nearest neighbor interaction with strength $J_{A(B)}$ between same-sublattice sites spaced by $\delta_z$ in the stacking direction $z$; $H_D$ is the intra-layer next nearest neighbor Dzyaloshinskii-Moriya (DM) interaction \cite{dzyaloshinskii1958, moriya1960} between same-sublattice sites connected by $\bm \beta_i$; and the last term $Z$ is the Zeeman energy. For simplicity, we choose the interaction strength as $D_A=-D_B=D>0$, $J_1(\bm \alpha_i) = J_1>0$, and $J_2(\bm \alpha_i \pm \delta_z \hat z) = J_2>0$. Moreover, since the Zeeman term only determines the spin polarization of the ground state, we henceforth set $B_z \rightarrow 0^+$. We further assume the inter-layer lattice constant $\delta_z=a$.

\begin{figure} [htb]
\includegraphics[width = 8cm]{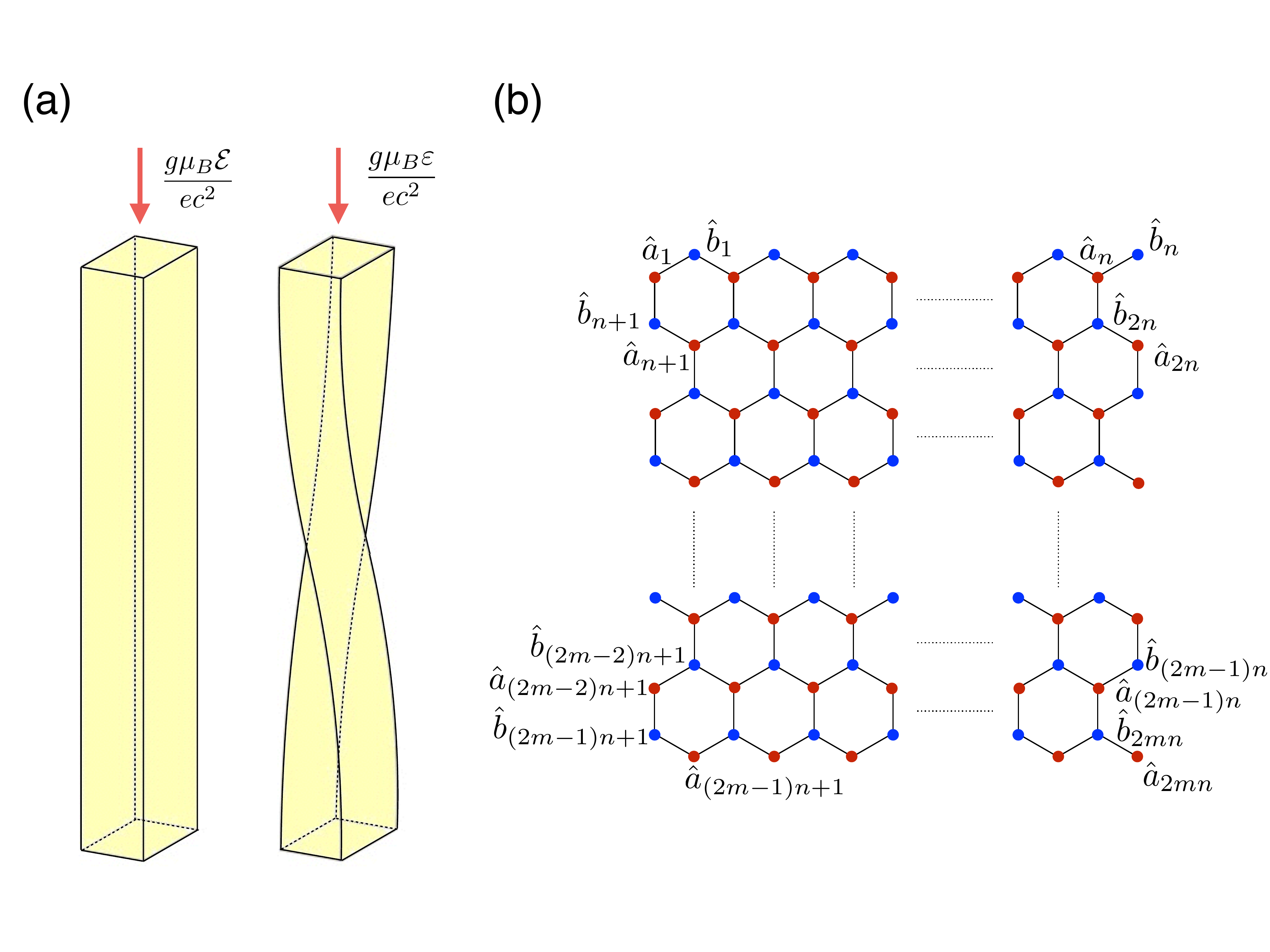}
\caption{Schematic plot for the Weyl ferromagnet nanowire. (a) Nanowire under an inhomogeneous electric field and nanowire under a twist deformation. Landau quantization takes place in both cases.  (b) Cross section of the Weyl ferromagnet nanowire with a pair of zigzag edges ($x$-direction) and a pair of armchair edges ($y$-direction). We use $m=n=30$ unless otherwise specified so that all numerical simulations could be implemented with available computational resources.} \label{fig_3}
\end{figure}

The Heisenberg Hamiltonian (Eq.~\ref{H_Heisenberg}) can be rewritten in terms of magnons through the Holstein-Primakoff transformation \cite{holstein1940}, 
\begin{align*}
S_\mu^+(\bm R) &=\sqrt{2S} \sqrt{1-\frac{\mu_{\bm R}^\dagger \mu_{\bm R}}{2S}} \mu_{\bm R},
\\
S_\mu^-(\bm R) &=\sqrt{2S} \mu_{\bm R}^\dagger \sqrt{1-\frac{\mu_{\bm R}^\dagger \mu_{\bm R}}{2S}}, 
\\
S_\mu^z(\bm R) &= S - \mu_{\bm R}^\dagger \mu_{\bm R},
\end{align*}
where $\bm R = (\bm R_\perp,z)$ indicates the position of the unit cell, $\mu=A,B$ is the sublattice index, and $\mu_{\bm R}^\dagger/\mu_{\bm R}$ is the corresponding magnon creation/annihilation operator. In the single-particle limit, the Hamiltonian becomes
\begin{equation}
H=E_\text{FM}+H_\text{M},
\end{equation}
where the ferromagnetic ground state energy $E_\text{FM} = -NS^2(K_A+K_B+J_A+J_B+3J_1+6J_2)$ and the tight binding magnon Hamiltonian is
\begin{multline}
H_\text{M}= 2K_A S \sum_{\bm R_\perp, z} a_{\bm R_\perp, z}^\dagger a_{\bm R_\perp, z}
 + 2K_B S \sum_{\bm R_\perp, z} b_{\bm R_\perp, z}^\dagger b_{\bm R_\perp, z}
 \\
+(3J_1+6J_2)S\sum_{\bm R_\perp, z}   (a_{\bm R_\perp, z}^\dagger a_{\bm R_\perp, z} + b_{\bm R_\perp, z}^\dagger b_{\bm R_\perp, z})
\\
+2J_A S\sum_{\bm R_\perp, z} a_{\bm R_\perp, z}^\dagger a_{\bm R_\perp, z}
+2J_B S \sum_{\bm R_\perp, z} b_{\bm R_\perp, z}^\dagger b_{\bm R_\perp, z}
\\
+\bigg\{ iDS \sum_{\bm R_\perp, z} \sum_{i} (-a_{\bm R_\perp, z}^\dagger a_{\bm R_\perp + \bm \beta_i, z} 
+b_{\bm R_\perp, z}^\dagger b_{\bm R_\perp + \bm \beta_i, z}) 
\\
-J_A S \sum_{\bm R_\perp, z} a_{\bm R_\perp, z}^\dagger a_{\bm R_\perp, z+a}
-J_B S \sum_{\bm R_\perp, z} b_{\bm R_\perp, z}^\dagger b_{\bm R_\perp, z+a} 
\\
-J_1S\sum_{\bm R_\perp, z} \sum_{i} a_{\bm R_\perp, z}^\dagger b_{\bm R_\perp + \bm \alpha_i - \bm \alpha_1, z}
\\ 
-J_2S \sum_{\bm R_\perp, z} \sum_{i} \sum_{s=\pm 1} a_{\bm R_\perp, z}^\dagger b_{\bm R_\perp + \bm \alpha_i - \bm \alpha_1, z +s a} + \text{H.c.} \bigg\}.
\end{multline}
We then perform Fourier transform
\begin{align}
a_{\bm R_\perp, z} = \frac{1}{\sqrt{N}} \sum_{\bm k} e^{i \bm k_\perp \cdot \bm R_\perp} e^{ik_zz} a_{\bm k},
\\
b_{\bm R_\perp, z} = \frac{1}{\sqrt{N}} \sum_{\bm k} e^{i \bm k_\perp \cdot (\bm R_\perp + \bm \alpha_1)} e^{ik_zz} b_{\bm k},
\end{align}
where $\bm k = (\bm k_\perp, k_z)$ and $N$ is the number of unit cells in the Weyl ferromagnet multilayer. Then the tight-binding Hamiltonian can be written as
\begin{equation}
H_\text{M} = \sum_{\bm k} \psi_{\bm k}^\dagger \pazocal{H}_{\bm k} \psi_{\bm k},
\end{equation}
where the sublattice basis is $\psi_{\bm k} = (a_{\bm k}, b_{\bm k})^T$ and the first-quantized Bloch Hamiltonian is
\begin{multline} \label{Bloch_Hk}
\pazocal{H}_{\bm k} = (-J_1 - 2J_2 \cos k_za) S \sum_{i} \cos(\bm k \cdot \bm \alpha_i) \sigma^x
\\ 
+ (J_1 + 2J_2 \cos k_za)  S\sum_{i} \sin(\bm k \cdot \bm \alpha_i) \sigma^y
\\
+ [K_- + J_-(1-\cos k_za) + 2D \sum_{i} \sin (\bm k \cdot \bm \beta_i) ]S \sigma^z
\\
+[3J_1 + 6J_2 +K_+ + J_+(1-\cos k_za)] S \sigma^0,
\end{multline}
where $\bm \sigma$ and $\sigma^0$ are Pauli matrices and the identity matrix, respectively, and we have used the notations $K_\pm = K_A \pm K_B$ and $J_\pm = J_A \pm J_B$. It is easy to find that $\pazocal{H}_{\bm k}$ can only be gapless at the corners of the 2D hexagonal Brillouin zone of the honeycomb ferromagnet. For simplicity, we require $K_-+3\sqrt{3}D>0$ but $-2J_1<K_- -3\sqrt{3}D<0$. In this case, there is only one pair of nodal points $\bm k_W^\eta=( -\frac{4\pi}{3\sqrt{3}a}, 0, \eta Q)$ with $\eta=\pm 1$ and
\begin{equation}
Q = \frac{1}{a}  \cos^{-1} \bigg( \frac{K_- + J_- - 3\sqrt{3}D}{J_-} \bigg). 
\end{equation}
In the vicinity of these Weyl points, the Bloch Hamiltonian can be expanded to the lowest order as
\begin{equation}
\pazocal{H}_{\bm k_W^\eta + \bm q} = \pazocal{H}_{\bm k_W^\eta} + \hbar v_0^\eta q_z \sigma^0 + \sum_{i=x,y,z} \hbar v_i^\eta q_i \sigma^i,
\end{equation}
with the corresponding velocity parameters defined as
\begin{align*}
v_x^\eta &= -\frac{3}{2\hbar} (J_1+2J_2\cos Qa) S a,
\\
v_y^\eta &= -\frac{3}{2\hbar} (J_1+2J_2\cos Qa) S a,
\\
v_z^\eta &= \frac{\eta}{\hbar}  J_- Sa \sin Qa, 
\\
v_0^\eta &= \frac{\eta}{\hbar}  J_+ Sa \sin Qa.
\end{align*}
Without loss of generality, we take $J_->0$. Then the Berry flux (in units of $\pi$) that flows into/out of the Weyl points is 
\begin{equation} \label{chirality}
\chi_\eta = \text{sgn} (v_xv_yv_z) =  \text{sgn} (v_z) =  \eta,
\end{equation}
which is locked to the momentum space position of Weyl points. The topological nature of $\pazocal{H}_{\bm k}$ can be verified by evaluating the Chern number of a 2D slice with fixed $k_z$, which is
\begin{equation} \label{Chern_number}
C_{k_z} = \frac{1}{2} \text{sgn}\big[K_- + J_- (1-\cos k_za )- 3\sqrt{3}D \big] - \frac{1}{2}.
\end{equation}
Based on the restrictions we impose on the parameters, we obtain
\begin{equation}
C_{k_z} = 
\begin{cases}
-1 & \mbox{$|k_z| < Q$}
\\
0 & \mbox{$Q<|k_z|<\pi$}
\end{cases}.
\end{equation}
\begin{figure*}[ht]
\includegraphics[width = 16cm]{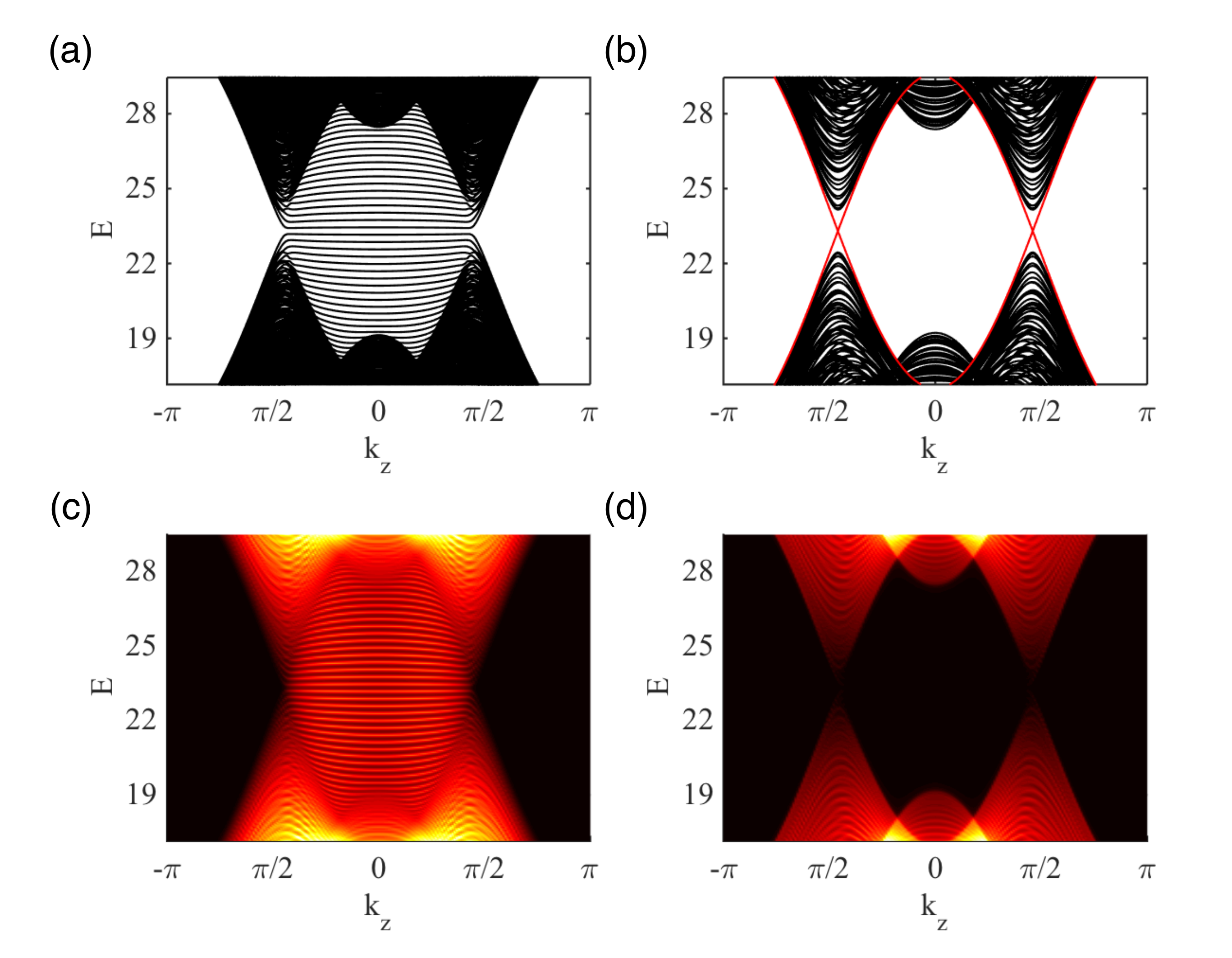}
\caption{Magnon dispersion and spectral functions for the Weyl ferromagnet multilayer. For all panels, we set $DS=1$ and measure energies in terms of $DS$ such that $J_1 S=4.56$, $J_2 S=1.14$, $J_- S=7.22$, $K_+ S=2.77$ and $K_- S=-1.12$. (a) Magnon band structure for the nanowire with a pair of zigzag edges and a pair of armchair edges in the cross section. The cross section of the nanowire is illustrated in Fig.~\ref{fig_3}(b). The magnon bands exhibit two Weyl points on the $k_z$ axis and are connected by a set of almost flat states analogous to the arc states in Weyl semimetals. (b) Magnon bands for the nanowire with periodic boundary conditions for the cross section. The flat bands disappear, indicating their surface origin. The red curves are the analytical dispersion $\epsilon_{\bm k} = [K_++3J_1+6J_2]S \pm [K_- + J_-(1-\cos k_za) - 3\sqrt{3}D]S$ for the Bloch Hamiltonian $\pazocal{H}_{\bm k}$ at the honeycomb lattice Brillouin zone corner $\bm k_\perp = (-4\pi/3\sqrt{3}a,0)$. (c) Surface spectral function of the Bloch Hamiltonian which confirms that the almost flat states reside on surfaces. (d) Bulk spectral function which indicates the positions of Weyl cones. } \label{fig_2}
\end{figure*}
Therefore, we expect arc surface states, which are akin to Fermi arcs in an electronic Weyl semimetal, connecting magnon Weyl points at the topological crystal momenta $|k_z|<Q$. To confirm this, we numerically calculate the band structure (Fig.~\ref{fig_2}(a)-(d)) of a Weyl ferromagnet nanowire whose cross section containing $1800$ lattice sites is schematically plotted in Fig.~\ref{fig_3}(b). The open boundary of the cross section consists of one pair of zigzag edges ($x$-direction) and one pair of armchair edges ($y$-direction). For simplicity and visual clarity, we set $J_+=0$ in Eq.~\ref{Bloch_Hk} while taking care to preserve the positive-definiteness of the magnon energy/spin wave frequency; the $J_+$ term will be reinstated in Appendix~\ref{a3}. As illustrated in Fig.~\ref{fig_2}(a), the Weyl ferromagnet nanowire exhibits a pair of Weyl cones connected by a set of almost flat bands. When the boundaries are closed, these flat bands disappear in Fig.~\ref{fig_2}(b), indicating that the flat bands only reside on the surface of the nanowire. This can be further confirmed by evaluating the spectral functions: these flat states have a large spectral density at the surface (Fig.~\ref{fig_2}(c)) but disappear deep in the bulk (Fig.~\ref{fig_2}(d)).

\section{Weyl ferromagnet under electromagnetic fields and strain}
\label{s3}
In Sec.~\ref{s2}, we introduced a multilayer model for the Weyl ferromagnet and discussed the magnon band topology. In order to obtain magnon quantum anomalies, gauge fields are needed to manipulate magnons. In the present section, we will first discuss the Landau quantization of magnon bands in the presence of an inhomogeneous electric field $\bm E$ (Sec.~\ref{s3a}) or an inhomogeneous chiral pseudo-electric field $\bm e^\eta$ (Sec.~\ref{s3b}) induced by a static twist deformation of the Weyl ferromagnet nanowire. Then we derive the equation of motion for magnons under an inhomogeneous magnetic field $\bm B$ (Sec.~\ref{s3c}) or an inhomogeneous chiral pseudo-magnetic field $\bm b^\eta$ (Sec.~\ref{s3d}) induced by a dynamic uniaxial strain.

\subsection{Landau quantization in the presence of $\bm E$}
\label{s3a}
To begin with, we study the magnon band structure under an electric field $\bm E$. Dual to the Aharonov-Bohm phase $\phi_{\text{AB}} = - \frac{e}{\hbar} \int_{\bm r}^{\bm r + \bm \delta} \bm A \cdot d \bm l$ acquired by electrons moving in a magnetic field $\bm B = \nabla \times \bm A$, magnons moving in an electric field can acquire an Aharonov-Casher phase \cite{aharonov1984}
\begin{equation}
\phi_{\text{AC}} = - \frac{e}{\hbar } \int_{\bm r}^{\bm r + \bm \delta} \frac{1}{ec^2} (\bm E \times \bm \mu) \cdot d \bm l,
\end{equation}
where the curl of the integrand $\frac{1}{e c^2} \nabla \times (\bm E \times \bm \mu)$ is dual to the magnetic field in electronics and will Landau-quantize the magnon bands. When transformed to the reciprocal space, the Aharonov-Casher phase results in the Peierls substitution $\bm k \rightarrow \bm k + \frac{e}{\hbar} \frac{1}{ec^2}\bm E \times \bm \mu$. Explicitly, if we take the magnon magnetic moment $\bm \mu = -g \mu_B \hat z$ and the electric field $\bm E = (\frac{1}{2} \pazocal{E} x, \frac{1}{2} \pazocal{E} y, 0)$, which may be experimentally realized by periodically arranging scanning tunneling microscope (STM) tips \cite{nakata2017, note4}. The resulting Dirac-Landau levels are given by
\begin{equation}
\epsilon_n^\eta(q_z) = \pm \hbar \sqrt{\big(v_z^\eta q_z\big)^2 + 2 n \Big|\frac{g \mu_B \pazocal{E}}{\hbar c^2} v_x^\eta v_y^\eta \Big|} \quad n=1,2,\dots
\end{equation}
and
\begin{multline}
\epsilon_0^\eta(q_z) = -\text{sgn}\bigg(\frac{g \mu_B \pazocal{E}}{\hbar c^2} v_x^\eta v_y^\eta \bigg) \hbar v_z^\eta q_z 
\\
=-\eta\text{sgn}(g\pazocal{E})\hbar|v_z^\eta|q_z.
\end{multline}
The higher ($n>1$) Landau levels at both Weyl points are identical and they always come in pairs with opposite energies at each momentum $q_z$. However, the zeroth Landau levels at the two Weyl points are not identical but counter-propagating. Without loss of generality, we choose $\sgn(g\pazocal{E})=-1$ such that the right (left) Weyl cone hosts a right (left) moving zeroth Landau level $\hbar|v_z^\eta|q_z$ ($-\hbar|v_z^\eta|q_z$). Unlike higher Landau levels, at each Weyl point, the zeroth Landau levels are unpaired.

\begin{figure*}[ht]
\includegraphics[width = 16cm]{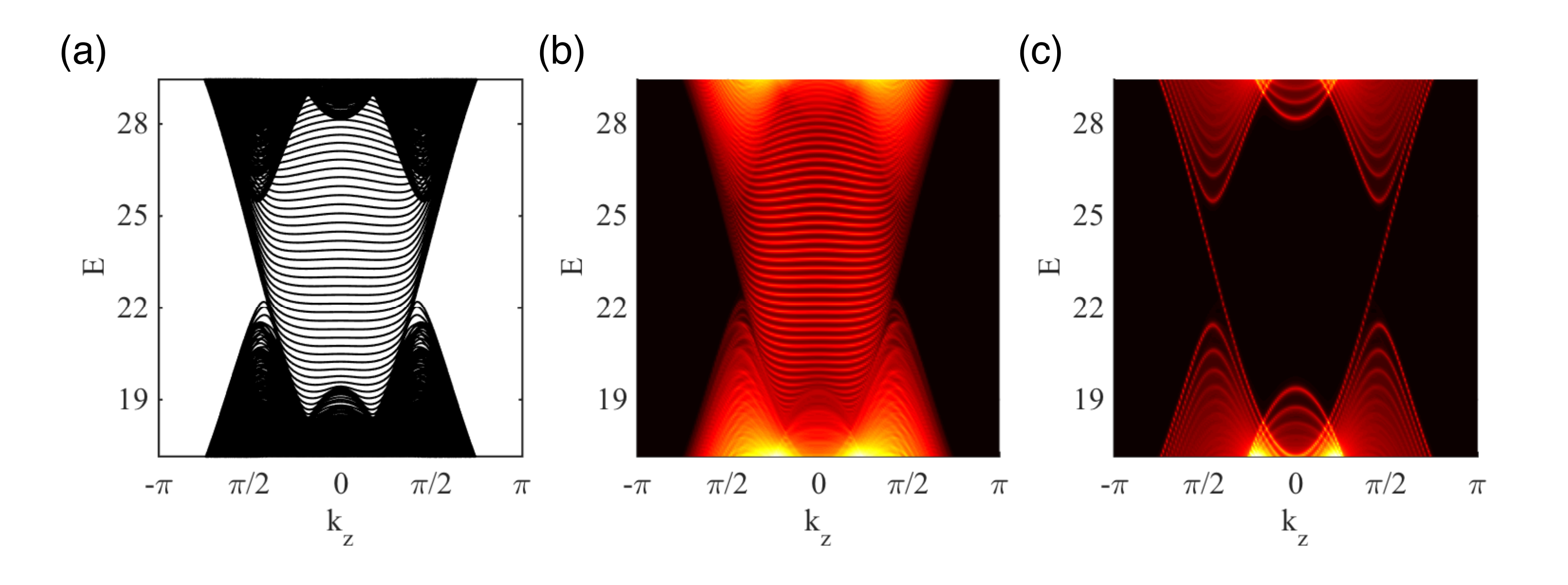}
\caption{Magnon dispersion of the Weyl ferromagnet nanowire under an inhomogeneous electric field. For all panels, we take $\frac{g\mu_B\pazocal{E}a^2}{ec^2}=-0.0124\Phi_0$ where $\pazocal E$ represents the gradient of the external electric field and $\Phi_0=h/2e$ is the magnetic flux quantum. (a) Magnon bands are Landau-quantized by the external inhomogeneous electric field due to the Aharonov-Casher effect. The two resulting zeroth Landau levels at different Weyl points have opposite velocities $\pm |v_z^\eta|$ and are connected by a set of almost flat states. (b) Surface spectral function, which reveals that these flat bands are localized at the surface of the Weyl ferromagnet nanowire. (c) Bulk spectral function highlighting the Dirac-Landau levels at each Weyl cone. } \label{fig_4}
\end{figure*}

To numerically verify the Landau quantization due to the Aharonov-Casher effect, we consider the nanowire geometry under an electric field (left panel, Fig.~\ref{fig_3}(a)). The calculated band structure of the Weyl ferromagnet nanowire under an inhomogeneous electric field is shown in Fig.~\ref{fig_4}(a). The zeroth Landau levels can be easily identified, and are connected by a set of almost flat states whose surface origin can be confirmed by calculating the surface spectral function (Fig.~\ref{fig_4}(b)). The obscured higher Landau levels, on the other hand, are revealed by the bulk spectral function as illustrated in Fig.~\ref{fig_4}(c).

\subsection{Landau quantization in the presence of $\bm e^\eta$}
\label{s3b}
In the context of Weyl semimetals, besides applying a magnetic field $\bm B$, the electronic bands are also Landau-quantized by a suitable lattice deformation \cite{liu2017a, sumiyoshi2016, pikulin2016, grushin2016, cortijo2015}, which spatially modulates the overlap integrals. The overall effect of such a deformation on the low-energy Hamiltonian is a minimal substitution analogous to that for a magnetic field. In other words, the lattice deformation due to external strain induces a pseudo-magnetic field. It is worth noting that such a strain-induced pseudo-magnetic field only couples to the Weyl points and becomes negligible at energies far away from the Weyl points. In this section, we will show that a similar strain-induced pseudo-electric field can be generated by spin lattice deformation and leads to Landau quantization of Weyl magnons.

\begin{figure}[htb]
\includegraphics[width = 7cm]{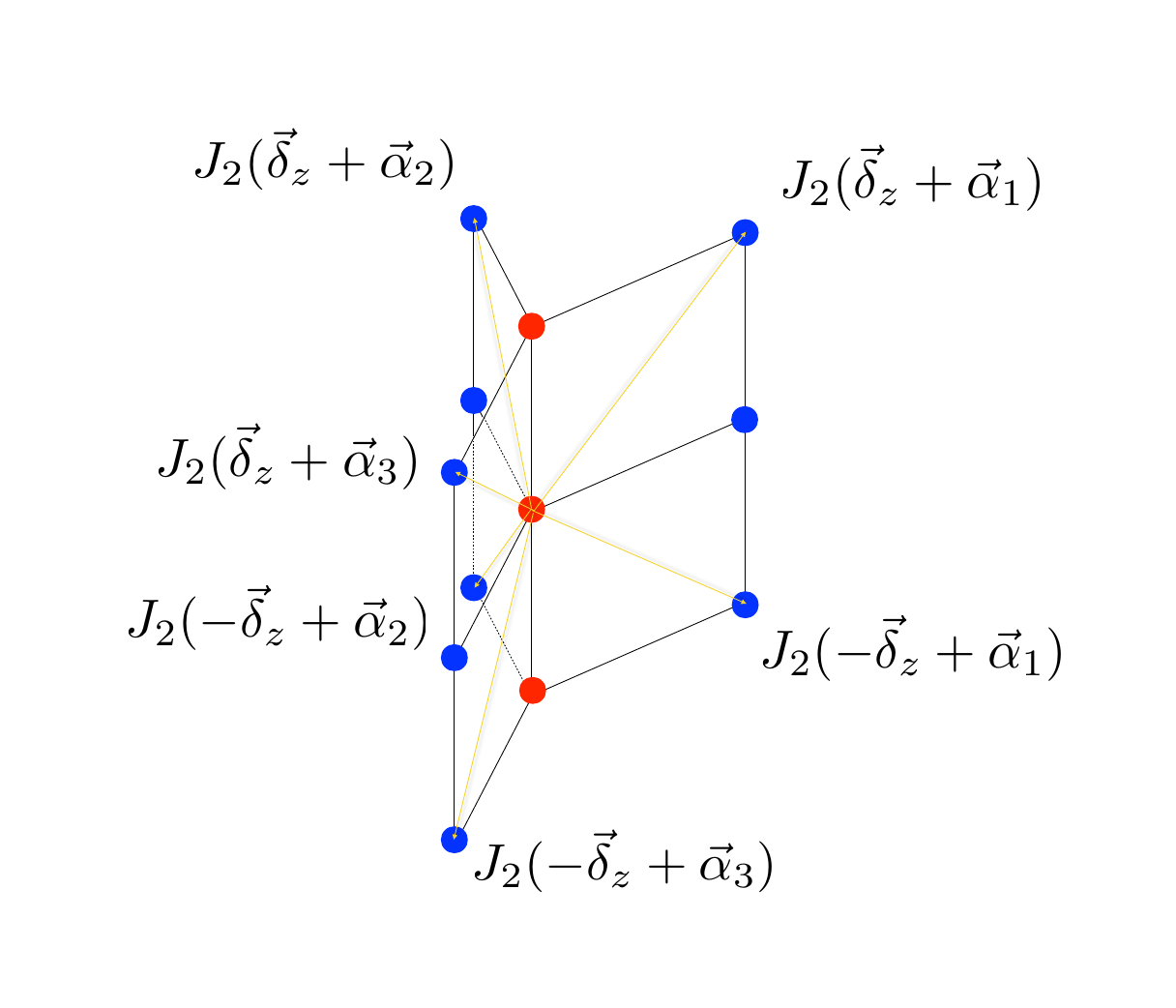}
\caption{Schematic plot for the exchange integrals $J_2$. The most important impact of the twist deformation is to modulate $J_2$ spatially.} \label{fig_5}
\end{figure}

\begin{figure*}[ht]
\includegraphics[width = 16cm]{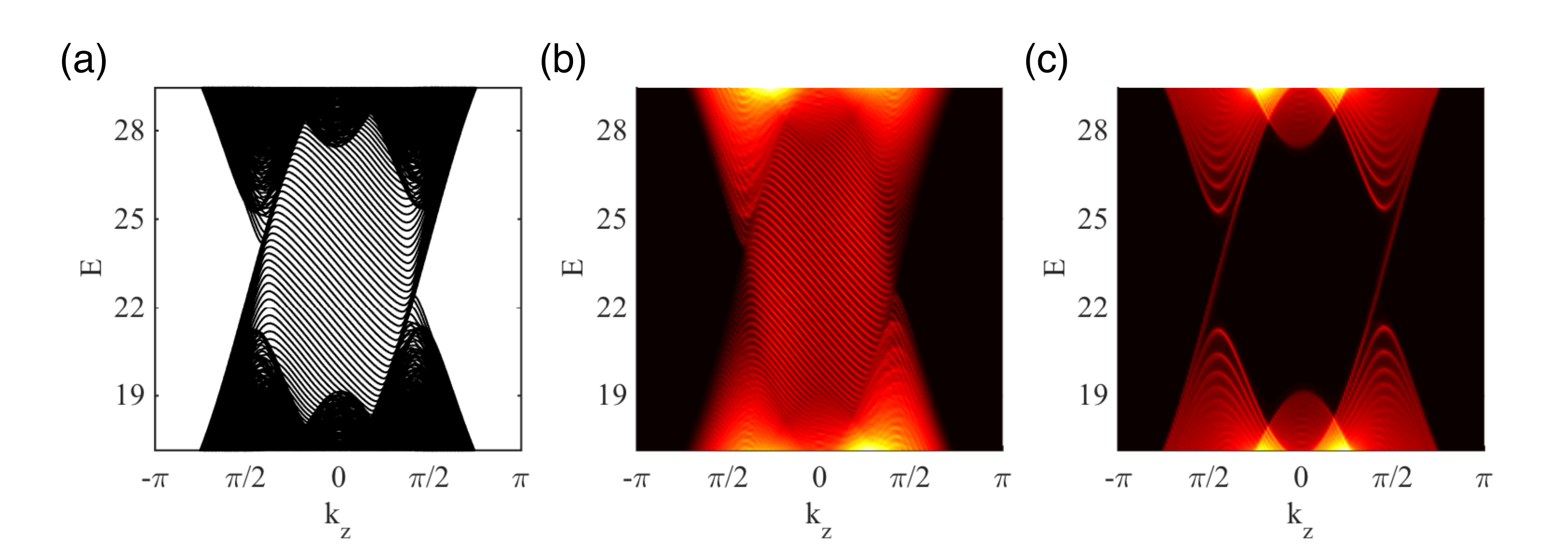}
\caption{Magnon dispersion of a twisted Weyl ferromagnet nanowire. For all panels, we take $\frac{g\mu_B \varepsilon a^2}{ec^2} =-0.0124\Phi_0$ where $\varepsilon$ represents the gradient of the strain-induced pseudo-electric field. (a) Magnon bands are Landau-quantized by the strain-induced pseudo-electric field. The resulting zeroth Landau levels at the two Weyl points are both right-moving and are connected by a set of left-moving states. (b) Surface spectral function, which reveals that these left-moving states are localized at the surface of the Weyl ferromagnet nanowire. (c) Bulk spectral function highlighting the Dirac-Landau levels at each Weyl cone. } \label{fig_6}
\end{figure*}

For concreteness, we consider the spin lattice deformation due to a twist (Fig.~\ref{fig_3}(a), right panel). The resulting displacement field is
\begin{equation}
\bm u = \frac{z}{L}(\bm \Omega \times \bm R_\perp),
\end{equation}
where $\Omega$ is the angle of rotation of the uppermost layer with respect to the lowermost layer, and $L$ is the spacing between these two layers, i.e., the length of the nanowire. For this displacement field, the non-zero components of the symmetric strain tensor $u_{ij} = \partial_i u_j + \partial_j u_i$ are $u_{13} = u_{31} = -\Omega y/2L $ and $ u_{23} = u_{32} = \Omega x/2L$. For this reason, we only need to consider the exchange integrals whose arguments simultaneously have out-of-plane and in-plane components (Appendix~\ref{a1}). In our model, this refers to the six $J_2$'s as illustrated in Fig.~\ref{fig_5}. The substitutions for these $J_2$'s are
\begin{align*}
J_2(\bm \alpha_1 \pm a \hat z) & \rightarrow J_2 ( 1\mp\tfrac{\sqrt{3}}{2}u_{31} \mp \tfrac{1}{2} u_{32} ),
\\
J_2(\bm \alpha_2 \pm a \hat z) & \rightarrow J_2 ( 1 \pm\tfrac{\sqrt{3}}{2}u_{31} \mp \tfrac{1}{2} u_{32} ),
\\
J_2(\bm \alpha_3 \pm a\hat z) & \rightarrow J_2 (1\pm u_{32})
\end{align*}
under which an additional term appears in $\pazocal{H}_{\bm k}$
\begin{equation} \label{twist_dH}
\delta \pazocal{H}_{\bm k}^{\bm e} = -2J_2S \sin k_za \sum_i d_i (\sin \bm k \cdot \bm \alpha_i \sigma^x + \cos \bm k \cdot \bm \alpha_i \sigma^y),
\end{equation}
where
\begin{equation*}
(d_1,d_2,d_3)= \Big( \frac{\sqrt{3}}{2}u_{31} + \frac{1}{2} u_{32}, -\frac{\sqrt{3}}{2}u_{31} + \frac{1}{2} u_{32}, -u_{32} \Big).
\end{equation*}
In the vicinity of the Weyl points $\bm k_W^\eta$, the Bloch Hamiltonian of the twisted Weyl ferromagnet nanowire reads
\begin{multline} \label{dirac_Hk1}
\pazocal{H}_{\bm k_W^\eta+\bm q} + \delta \pazocal{H}_{\bm k_W^\eta+\bm q}^{\bm e} \approx \pazocal{H}_{\bm k_W^\eta} + \hbar v_0^\eta q_z \sigma^0 + \sum_i \hbar v_i^\eta q_i \sigma^i + \delta \pazocal{H}_{\bm k_W^\eta}^{\bm e}
\\
= \pazocal{H}_{\bm k_W^\eta} + \hbar v_0^\eta \Big(q_z + \frac{e}{\hbar} a_{S,z}^\eta \Big) \sigma^0 + \sum_i \hbar v_i^\eta \Big(q_i + \frac{e}{\hbar} a_{S,i}^\eta \Big) \sigma^i.
\end{multline}
It is worth noting that the strain-induced term can be incorporated into the linearized Hamiltonian through a minimal substitution $\bm q \rightarrow \bm q +\frac{e}{\hbar} \bm a_S^\eta$, where the strain-induced vector potential is
\begin{equation} \label{vp_static}
\bm a_S^\eta= -\eta \frac{2\hbar}{ea}\frac{J_2 \sin Qa}{J_1 + 2J_2 \cos Qa} (u_{31}, u_{32}, 0),
\end{equation}
which is a chiral gauge field taking opposite values at different Weyl points. We note that the multilayer model in Ref.~\cite{su2017a} does not have such a strain-induced vector potential because the inter-layer next nearest neighbor interaction is not considered, i.e., $J_2=0$. As $\nabla \times \bm a_S^\eta \neq 0$, this strain-induced vector potential will result in Landau quantization of magnon bands. In Sec.~\ref{s3a}, we demonstrate that electric field $\bm E = (\frac{1}{2} \pazocal E x, \frac{1}{2} \pazocal E y, 0)$ produces Landau quantization through the Aharonov-Casher effect. Therefore, we may interpret $\bm a_S^\eta$ as the vector potential of a chiral pseudo-electric field $\bm e^\eta = \eta \bm e =\eta (\frac{1}{2} \varepsilon x, \frac{1}{2} \varepsilon y, 0)$, which only differs from $\bm E$ by a chiral charge $\eta$. The field gradient of this pseudo-electric field can be determined by $\nabla \times \bm a_S^\eta = \frac{1}{ec^2} \nabla \times (\bm e^\eta \times \bm \mu) = \eta \frac{g\mu_B \varepsilon}{ec^2} \hat z $. Explicitly, the gradient reads
\begin{equation}
\varepsilon = -\frac{2\hbar}{ea}\frac{J_2 \sin Qa}{J_1 + 2J_2 \cos Qa} \frac{\Omega}{L} \frac{ec^2}{g\mu_B}.
\end{equation}
It is critically important to note that the strain-induced vector potential is a unique feature of relativistic particles. We thus expect that the strain-induced pseudo-electric field and magnon Dirac-Landau levels only reside in the vicinity of magnon Weyl points. Unlike the electric field $\bm E$ which produces counter-propagating zeroth Landau levels, the chiral pseudo-electric field $\bm e^\eta$ results in co-propagating zeroth Landau levels as illustrated in Fig.~\ref{fig_6}(a,c). At the first glance, this causes an imbalance in the numbers of left-moving and right-moving channels, which must be the same in a lattice model. Considering that we have only discussed bulk Landau levels so far, we argue that there should be a set of surface states connecting the bulk Landau levels at the two Weyl cones, providing the needed counter-propagating channels to compensate the imbalance. Our argument has been confirmed by the numerical simulation of the surface spectral function as illustrated in Fig.~\ref{fig_6}(b). Remarkably, we note that inside the gap spanned by the two first Landau levels, the left-moving surface channels and right-moving bulk channels are spatially well separated and cannot be scattered into each other. Therefore, if the reservoirs are fine-tuned to populate this gap with magnons \cite{su2017a, meier2003}, the bulk magnon particle current and surface magnon particle current are counter-propagating in the ballistic regime. This leads to exotic spin and heat transport known as the bulk-surface separation which will be elaborated in Sec.~\ref{s4}.

\subsection{Magnon motion in the presence of $\bm B$}
\label{s3c}
In Sec.~\ref{s3a}, we have Landau-quantized the magnon bands by applying an inhomogeneous electric field $\bm E$. To implement quantum anomalies with magnons, we need to pump them through the zeroth Landau levels. In this section, we will elucidate how magnons are pumped by an inhomogeneous magnetic field $\bm B$. The equation of motion of magnons will be derived.

Due to the magnetic moment it carries, a magnon has a Zeeman energy $U=-\bm \mu \cdot \bm B $ in the presence of a magnetic field $\bm B$. The force exerted on the magnon is then $\bm F= -\nabla U =\nabla(\bm \mu \cdot \bm B) = \hbar \frac{d\bm k}{dt}$. Practically, we are only concerned about the transport in the $z$ direction, which is governed by
\begin{equation} \label{EOM_Bz}
\hbar \frac{d k_z}{dt} = \partial_z (\bm \mu \cdot \bm B).
\end{equation}
Therefore, an inhomogeneous magnetic field in the $z$-direction will be capable of driving magnons from one magnon Weyl point to the other. More generally, we consider the case where there is also an electric field $\bm E$ in addition to $\bm B$. As discussed in Sec.~\ref{s3a}, due to the Aharonov-Casher phase, the canonical momentum will be shifted through the Peierls substitution $\bm p \rightarrow \bm p + \frac{1}{c^2} \bm E \times \bm \mu$ in the presence of $\bm E$. Then the magnon Hamiltonian can be written down as
\begin{equation}
\mathcal H_M = K\big(\bm p + \tfrac{1}{c^2} \bm E \times \bm \mu \big) - \bm \mu \cdot \bm B,
\end{equation}
where $K(\bm p)$ is the magnon kinetic energy whose specific form does not matter for our purpose. The Hamilton's equations of motion give
\begin{equation}
\bm v = \frac{\partial \mathcal H_M}{\partial \bm p} = \frac{\partial K}{\partial \bm p}
\end{equation}
and
\begin{equation}
\frac{d \bm p}{dt} = -\frac{\partial \mathcal H_M}{\partial \bm r} = \frac{1}{c^2} \nabla[\bm v \cdot (\bm \mu \times \bm E)] + \nabla(\bm \mu \cdot \bm B).
\end{equation}
Therefore, the equation of motion for the crystal momentum $\bm k = \frac{1}{\hbar}\bm p + \frac{1}{\hbar c^2} \bm E \times \bm \mu$ is
\begin{equation} \label{EOM_BE}
\hbar \frac{d \bm k}{dt} = \nabla(\bm \mu \cdot \bm B) -\frac{\partial}{\partial t} \frac{1}{c^2} \bm \mu \times \bm E + \frac{1}{c^2} \bm v \times [\nabla \times (\bm \mu \times \bm E)],
\end{equation}
which clearly shows the duality to electronics with the magnon Zeeman energy $-\bm \mu\cdot \bm B$ (the electric momentum $\bm \mu\times \bm E$) playing the role of the electric potential energy $-e\phi$ (the magnetic momentum $e\bm A$). For the inhomogeneous electric field $\bm E = (\frac{1}{2} \pazocal{E}x, \frac{1}{2} \pazocal{E} y, 0)$ and the magnetic moment $\bm \mu = -g\mu_B \hat z$ used in Sec.~\ref{s3a}, the $z$ component of Eq.~\ref{EOM_BE} is exactly Eq.~\ref{EOM_Bz}.

In summary, an inhomogeneous magnetic field can be used to pump magnons along the $z$ direction and is capable of producing magnon quantum anomalies. This will be discussed in detail in Sec.~\ref{s4a} and Sec.~\ref{s4d}. In the meanwhile, a static inhomogeneous electric field only provides Landau quantization and does not affect the magnon transport in the $z$ direction.

\subsection{Magnon motion in the presence of $\bm b^\eta$}
\label{s3d}
In Sec.~\ref{s3b}, we have shown that magnon bands can be Landau-quantized by a chiral pseudo-electric field $\bm e^\eta$ induced by a static twist, under which the magnon Hamiltonian is modified by the Peierls substitution $\bm q \rightarrow \bm q + \frac{e}{\hbar} \frac{1}{ec^2} \bm e^\eta \times \bm \mu$. Therefore, by comparing to Eq.~\ref{EOM_BE}, the magnon equation of motion in the presence of $\bm e^\eta$ and $\bm B$ can be immediately written down as
\begin{equation} \label{EOM_Be}
\hbar \frac{d \bm k}{dt} = \nabla(\bm \mu \cdot \bm B) -\frac{\partial}{\partial t} \frac{1}{c^2} \bm \mu \times \bm e^\eta + \frac{1}{c^2} \bm v \times [\nabla \times (\bm \mu \times \bm e^\eta)].
\end{equation}
Because $\bm e^\eta$ field has exactly the same spacetime dependence as the inhomogeneous electric field $\bm E$ except for the chiral nature, our analysis for $\bm E$ in Sec.~\ref{s3c} can be transplanted to $\bm e^\eta$. Therefore we argue that magnons cannot be pumped by the pseudo-electric field induced by a static twist deformation. For this reason, in order to pump magnons, we resort to a $\textit{dynamic}$ deformation.

We consider a dynamic uniaxial strain whose only nonzero strain tensor component is $u_{33}$. Such a strain can be generated by the displacement field $\bm u = u_z(z,t) \hat z$. The knowledge of the explicit form of $u_z$ is not necessary for our purpose. Experimentally, a legitimate $u_z$ may be obtained by applying ultrasonic sound waves along the $z$ direction. Under this uniaxial strain $u_{33}$, we need to modify the exchange integrals whose arguments has nonzero $z$ components (Appendix~\ref{a1}). Such exchange integrals are $J_A$, $J_B$, and the six $J_2$'s illustrated in Fig.~\ref{fig_5}. The substitutions are
\begin{align*}
J_A \rightarrow J_A(1-u_{33})&
\\
J_B \rightarrow J_B(1-u_{33})& ,
\\
J_2 \rightarrow J_2(1- \tfrac{1}{2}u_{33})&
\end{align*}
under which an extra term enters the Bloch Hamiltonian (Eq.~\ref{Bloch_Hk}),
\begin{multline} \label{uniaxial_dH}
\delta \pazocal{H}_{\bm k}^{\bm b} =
u_{33}J_2 S \cos k_za \sum_i \cos (\bm k \cdot \bm \alpha_i) \sigma^x
\\
-u_{33}J_2 S \cos k_za \sum_i \sin (\bm k \cdot \bm \alpha_i)\sigma^y
\\
- u_{33}J_- S (1-\cos k_za) \sigma^z 
\\
- u_{33}[3 J_2  + J_+ (1- \cos k_za)]S \sigma^0.
\end{multline}
In the vicinity of the Weyl points $\bm k_W^\eta$, the Bloch Hamiltonian under the uniaxial deformation becomes
\begin{multline} \label{dirac_Hk2}
\pazocal{H}_{\bm k_W^\eta + \bm q} + \delta \pazocal{H}_{\bm k_W^\eta + \bm q}^{\bm b} \approx \pazocal{H}_{\bm k_W^\eta} + \hbar v_0^\eta \Big(q_z+\frac{e}{\hbar}a_{D,z}^\eta\Big)\sigma^0 \\+ \sum_i \hbar v_i^\eta \Big(q_i+\frac{e}{\hbar}a_{D,i}^\eta\Big)\sigma^i -3J_2Su_{33}\sigma^0.
\end{multline}
Similar to the static twist (Eq.~\ref{dirac_Hk1}), the dynamic uniaxial strain also shifts the momentum through the minimal substitution $\bm q \rightarrow \bm q + \frac{e}{\hbar} \bm a_D^\eta$, where the strain-induced vector potential reads
\begin{equation} \label{vp_dynamic}
\bm a_D^\eta = -\eta \frac{\hbar}{ea} \frac{1-\cos Qa}{\sin Qa}  (0,0,u_{33}).
\end{equation}
Similar to the its static counterpart $\bm a_S^\eta$, this dynamic strain-induced vector potential $\bm a_D^\eta$ is also chiral. However, it is critically important to note that this vector potential cannot be interpreted as the vector potential of a pseudo-electric field because $\nabla \times \bm a_D^\eta = 0$. In fact, $\bm a_D^\eta$ is associated with a chiral pseudo-magnetic field. To see this, by replicating the derivation in Sec.~\ref{s3c}, we first write down the equation of motion for magnons under the dynamic uniaxial strain
\begin{equation} \label{EOM_bB}
\hbar \frac{d\bm k}{dt} = e\frac{\partial \bm a_D^\eta}{\partial t} -e \bm v \times (\nabla \times \bm a_D^\eta) + \nabla(3J_2Su_{33}).
\end{equation} 
The extra non-chiral gradient term $\nabla(3J_2Su_{33})$ appears because the dynamic strain induces in the Hamiltonian an on-site term $3J_2Su_{33}\sigma^0$, which cannot be characterized by the minimal substitution. However, in systems where $J_2$ is reasonably small, namely $J_2S \ll \hbar c_s /a$ ($c_s$ is the speed of sound in the nanowire), this non-chiral term becomes negligible relative to the chiral time-derivative term \cite{note3}. We concentrate on such systems in the remainder of this paper. Therefore, considering the fact that $\nabla \times \bm a_D^\eta=0$, the $z$ component of Eq.~\ref{EOM_bB} is reduced to
\begin{equation} \label{EOM_bz}
\hbar \frac{dk_z}{dt} = e \frac{\partial a_{D,z}^\eta}{\partial t} = \partial_z(\bm \mu \cdot \bm b^\eta),
\end{equation}
where
\begin{equation} \label{ps_b}
\bm b^\eta= \frac{\bm \mu}{\mu^2} \int dz e \frac{\partial a_{D,z}^\eta}{\partial t} = \frac{\eta}{g\mu_B}\frac{\hbar}{a} \frac{1-\cos Qa}{\sin Qa} \int dz \frac{\partial u_{33}}{\partial t} \hat z
\end{equation}
can be understood as an inhomogeneous pseudo-magnetic field. Due to its chiral nature, $\bm b^\eta$ pumps magnons oppositely in the vicinity of the two Weyl points.

To sum up, a dynamic uniaxial strain can induce a chiral pseudo-magnetic field, accompanied by a non-chiral field whose effects can be neglected for reasonably small $J_2$. The pseudo-magnetic field pumps magnons in opposite directions $\pm z$ at the two Weyl points. Therefore, it may also result in magnon quantum anomalies as will be detailed in Sec.~\ref{s4b} and Sec.~\ref{s4c}. In the meanwhile, the static torsional strain only provides the needed Landau quantization and does not affect the magnon transport in the $z$ direction.

\section{Magnon quantum anomalies and the anomalous transport}
\label{s4}
In Sec.~\ref{s3}, we have found that magnon bands can be Landau-quantized by either an inhomogeneous transverse electric field $\bm E$ or a chiral pseudo-electric field $\bm e^\eta$ induced by a static twist. To drive the magnons along the zeroth Landau levels, we may use either an inhomogeneous longitudinal magnetic field $\bm B$, or a chiral pseudo-magnetic field $\bm b^\eta$ induced by a dynamic uniaxial strain. In this section, we will show that each of the four possible combinations of an electric/pseudo-electric field and a magnetic/pseudo-magnetic field gives rise to a magnon quantum anomaly. We will derive the anomaly equations and discuss the associated anomalous spin and heat transport.

\subsection{Magnon chiral anomaly due to $\bm E$ and $\bm B$}
\label{s4a}
In the present section, for completeness, we will derive the magnon chiral anomaly and the associated anomalous spin and heat currents in the presence of $\bm E$ and $\bm B$, though these have been briefly mentioned in Ref.~\cite{su2017a}. 

\begin{figure*}[htb]
\includegraphics[width = 16cm]{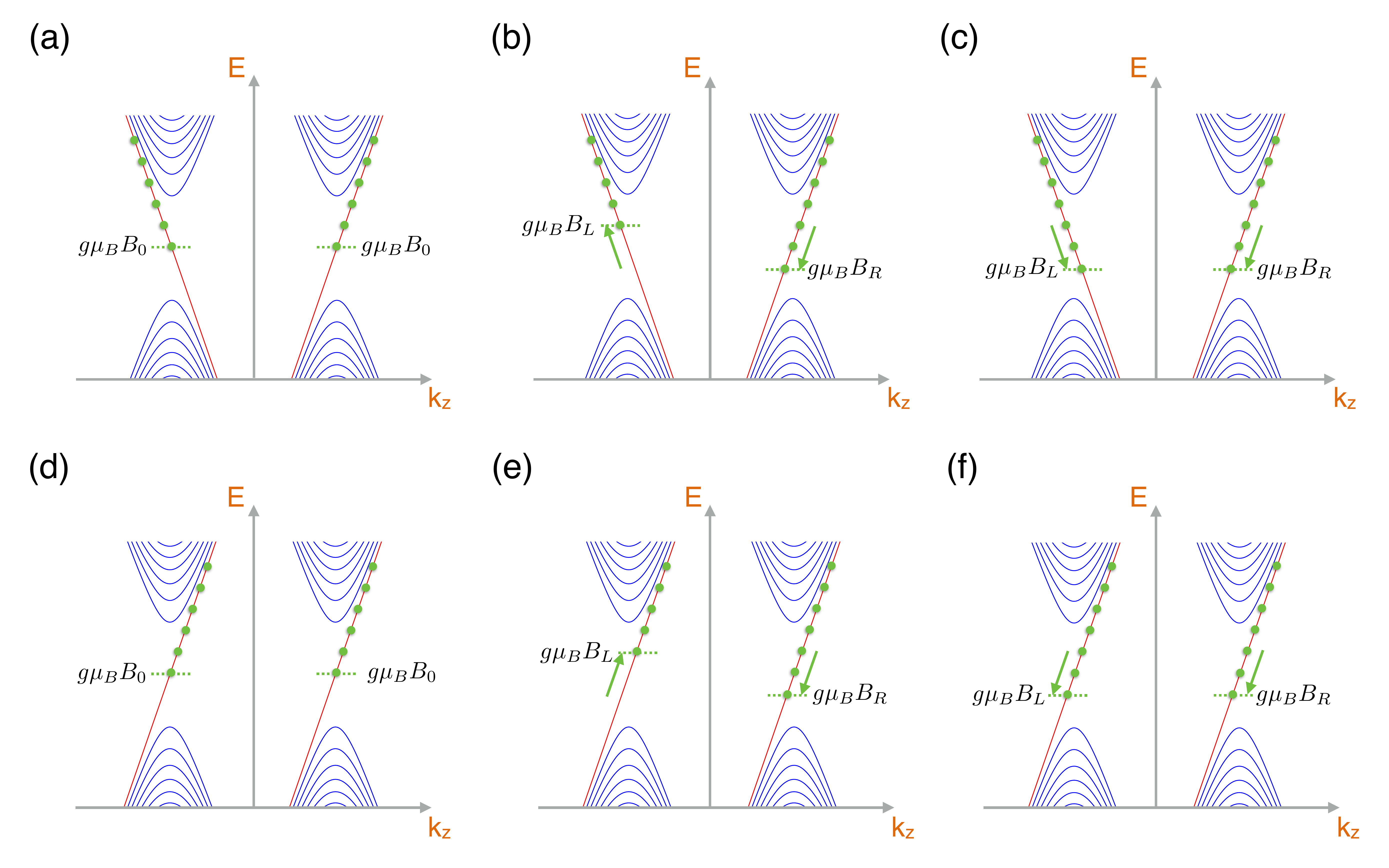}
\caption{Schematic plot of the magnon band structures and distributions in various quantum anomalies of a Weyl ferromagnet, which is in contact with two magnon reservoirs in a uniform magnetic field $\bm B_0$. (a)-(c) Magnon Dirac-Landau levels due to an inhomogeneous electric field. (d)-(f) Magnon Dirac-Landau levels due to a strain-induced pseudo-electric field. (a, d) Magnon distributions in the absence of pumping. (b, e) Magnon chiral anomaly with chirality imbalance created by ordinary magnetic field pumping in (b) and by pseudo-magnetic field pumping in (e). (c, f) Magnon heat anomaly with magnon concentration variation created by pseudo-magnetic field pumping in (c) and by ordinary magnetic field pumping in (f). For all panels, only the distributions (green dots) on the zeroth Landau levels (red) are plotted. In principle, magnons can occupy all bands above the population edges provided that the relaxation time is sufficiently long.} \label{fig_7}
\end{figure*}

We consider a Weyl ferromagnet nanowire aligned in the $z$ direction under the electric field $\bm E = (\frac{1}{2} \pazocal {E} x, \frac{1}{2} \pazocal{E} y, 0)$, whose magnon Landau levels are illustrated in Fig.~\ref{fig_4}. The left and right ends are attached to magnon reservoirs subjected to a uniform magnetic field $\bm B_0 = B_0 \hat z$.  Magnons will then attain a Zeeman energy $-\bm \mu \cdot \bm B_0 = g\mu_B B_0$ and the magnon population edge, which is originally located in the band minima, can be lifted up to the gap of the first Landau levels (Fig.~\ref{fig_7}(a)) by a suitable $B_0$. This is an analog to putting the chemical potential in the gap of Landau levels in Weyl semimetals \cite{su2017a, meier2003}. While electrons contributing to ballistic transport can have energies above or below the chemical potential, magnons must reside above the population edge $g\mu_B B_0$ in the magnon reservoirs to participate in ballistic transport. Then a spatially varying magnetic field $\bm B=B_z\hat z$ is overlaid, where $B_z$ has a nonzero gradient $\partial_zB_z= \pazocal{B}$. Based on Eq.~\ref{EOM_Bz}, the magnons begin to propagate along the Landau levels in the $-z$ direction according to semiclassical equation of motion $q_z(t)=q_z(0)-g\mu_B\int_0^t \pazocal{B} dt'/\hbar$. Thus the magnons originally on the left zeroth Landau level will be pumped to the right zeroth Landau level across the Brillouin zone boundary, which results in a chirality imbalance, a key feature of the magnon chiral anomaly. 

During this pumping process, the magnon population edge is gradually elevated (lowered) on the left (right) zeroth Landau level (Fig.~\ref{fig_7}(b)). The difference between the left and right magnon population edges is analogous to the chiral chemical potential in Weyl semimetals. Here we will see the difference in magnon population edges as a magnetic field bias $B_5 \ll B_0$ such that the left (right) Weyl cone experiences a magnetic field $B_L=B_0 + B_5$ ($B_R = B_0-B_5$). From the semiclassical equation of motion, we obtain
\begin{equation}
B_5 = -\frac{\hbar |v_z^\eta| \int dq_z}{g\mu_B} = \int_0^t \pazocal{B} |v_z^\eta| dt'.
\end{equation}
We now derive the chiral anomaly equation for magnons. The magnon concentration variation on the right zeroth Landau level can be written down as a Taylor series
\begin{multline} \label{nR_EB}
n_R^{\bm E, \bm B} = \int_{g\mu_B B_R}^{g\mu_B B_0} g_{\bm E}(\epsilon) n_B(\epsilon) d\epsilon 
= \frac{n_B(g\mu_B B_0)}{4\pi^2 l_E^2} \frac{g\mu_B B_5}{\hbar |v_z^\eta|} 
\\
- \sum_{n=1} \frac{n_B^{(n)}(g\mu_B B_0)}{(n+1)!} \frac{(-g \mu_B B_5)^{n+1}}{4\pi^2 l_E^2 \hbar |v_z^\eta|},  
\end{multline}
where $g_{\bm E}(\epsilon) = \frac{1}{2\pi l_E^2} \frac{1}{2\pi\hbar |v_z^\eta|}$ is the density of states with the electric length $l_E = (-\hbar c^2/g\mu_B\pazocal{E})^{1/2}$ \cite{note1}, and $n_B(\epsilon) = (e^{\epsilon/k_BT}-1)^{-1}$ is the magnon distribution function. Similarly, the magnon concentration variation on the left zeroth Landau level is
\begin{multline} \label{nL_EB}
n_L^{\bm E, \bm B} = \int_{g\mu_BB_L}^{g\mu_BB_0}g_{\bm E}(\epsilon) n_B(\epsilon) d\epsilon = - \frac{n_B(g\mu_B B_0)}{4\pi^2 l_E^2} \frac{g\mu_B B_5}{\hbar |v_z^\eta|} 
\\
- \sum_{n=1} \frac{n_B^{(n)}(g\mu_B B_0)}{(n+1)!} \frac{(g\mu_B B_5)^{n+1}}{ 4\pi^2 l_E^2 \hbar |v_z^\eta|}.
\end{multline}
In the low bias limit $g\mu_BB_5 \ll k_BT$, the net chirality pumping rate can be well approximated as
\begin{multline} \label{QA_chiral_EB}
\frac{d\rho_5^{\bm E, \bm B}}{dt} = \chi_R \frac{dn_R^{\bm E, \bm B}}{dt} + \chi_L \frac{dn_L^{\bm E, \bm B}}{dt} 
\\
\approx - \frac{g^2\mu_B^2}{2\pi^2\hbar^2c^2}n_B(g \mu_B B_0) \pazocal{EB},
\end{multline}
where, as in Eq.~\ref{chirality}, $\chi_R=+1$ and $\chi_L=-1$. More generally, for arbitrarily oriented $\bm E$ and $\bm B$, the magnon chiral anomaly equation reads
\begin{equation} \label{QA_EB}
\frac{d\rho_5^{\bm E, \bm B}}{dt} + \nabla \cdot \bm j_5^{\bm E,\bm B} \approx  \frac{n_B(g \mu_B B_0)}{2\pi^2\hbar^2c^2} \nabla(\bm \mu \cdot \bm B) \cdot [\nabla \times (\bm E \times \bm \mu)].
\end{equation}
Such a magnon chiral anomaly is dual to the electron chiral anomaly \cite{adler1969, bell1969, nielsen1983}. It is worth noting that Eq.~\ref{QA_EB} is true to the first order in $B_5$; at this order the number of magnons pumped out of the left zeroth Landau level is approximately equal to the number of magnons pumped into the right zeroth Landau level $-n_L \approx n_R$. However, at the second order in $B_5$, we have
\begin{multline}
n_R^{\bm E, \bm B}+n_L^{\bm E, \bm B} \\= -  \sum_{n=1} \frac{n_B^{(n)}(g\mu_B B_0)}{(n+1)!} \frac{ (g \mu_B B_5)^{n+1} [1+(-1)^{n+1}] }{4\pi^2 l_E^2 \hbar |v_z^\eta|}
\\ \approx
\frac{e^{g\mu_B B_0/k_BT}}{(e^{g\mu_B B_0/k_BT}-1)^2} \frac{1}{k_BT} \frac{(g\mu_B B_5)^2}{{4\pi^2 l_E^2 \hbar |v_z^\eta|}} >0,
\end{multline}
indicating that the magnon number is not conserved, because magnons are bosonic collective excitations rather than fundamental particles. From another point of view, if the magnon number were conserved, higher-energy magnons pumped out of the left zeroth Landau level would result in the same number of lower-energy magnons populating the right zeroth Landau level, and the magnon distribution in Fig.~\ref{fig_7}(b) would have a lower total energy than that in Fig.~\ref{fig_7}(a), making the pumping process spontaneous. However, magnon pumping actually requires energy injection by EM fields. Explicitly, the energy injection into the right zeroth Landau level is
\begin{multline} \label{UR_EB}
U_R^{\bm E, \bm B}  = \int_{g\mu_B B_R}^{+\infty} \epsilon g_{\bm E}(\epsilon) n_B(\epsilon) d\epsilon - \int_{g\mu_B B_0}^{+\infty} \epsilon g_{\bm E}(\epsilon) n_B(\epsilon) d\epsilon 
\\ 
= \frac{1}{4\pi^2 l_E^2 \hbar |v_z^\eta|} \bigg\{ \frac{1}{2} n_B(g\mu_B B_0) g^2\mu_B^2 (2B_0B_5 - B_5^2) 
\\
- \sum_{n=1} \frac{n_B^{(n)}(g\mu_B B_0)}{n!} \frac{g\mu_B B_0  ( - g\mu_B B_5)^{n+1}}{n+1}
\\
- \sum_{n=1} \frac{n_B^{(n)}(g\mu_B B_0)}{n!} \frac{ (- g\mu_B B_5)^{n+2}}{n+2} \bigg\}
\\ 
\approx \frac{1}{4\pi^2 l_E^2 \hbar |v_z^\eta|} \bigg\{ \frac{1}{2} n_B(g\mu_B B_0) g^2\mu_B^2 (2B_0B_5 - B_5^2) 
\\
- n_B'(g\mu_B B_0) \frac{g\mu_B B_0 (-g\mu_B B_5)^2 }{2}  \bigg\},
\end{multline}
where the approximation is still taken in the low bias limit $g\mu_BB_5 \ll k_BT$. Similarly, the energy depletion on the left zeroth Landau level can be easily written down by making the substitution $-B_5 \rightarrow +B_5$,
\begin{multline} \label{UL_EB}
U_L^{\bm E, \bm B} =  \int_{g\mu_B B_L}^{+\infty} \epsilon g_{\bm E}(\epsilon) n_B(\epsilon) d\epsilon - \int_{g\mu_B B_0}^{+\infty} \epsilon g_{\bm E}(\epsilon) n_B(\epsilon) d\epsilon 
\\
\approx \frac{1}{4\pi^2 l_E^2 \hbar |v_z^\eta|} \bigg\{ \frac{1}{2} n_B(g\mu_B B_0) g^2\mu_B^2 (- 2B_0B_5-B_5^2)
\\
- n_B'(g\mu_B B_0) \frac{g\mu_B B_0 (g\mu_B B_5)^2 }{2} \bigg\}.
\end{multline}
And the total energy variation
\begin{multline}
U_R^{\bm E, \bm B} + U_L^{\bm E, \bm B} = - \frac{1}{4\pi^2 l_E^2 \hbar |v_z^\eta|} n_B(g\mu_B B_0) (g\mu_B B_5)^2 
\\
- \frac{1}{4\pi^2 l_E^2 \hbar |v_z^\eta|} n_B'(g\mu_B B_0) g\mu_B B_0 (g\mu_B B_5)^2
\\
= - \frac{(g\mu_B B_5)^2}{4\pi^2 l_E^2 \hbar |v_z^\eta|} \frac{d}{d \epsilon} \epsilon n_B(\epsilon) \bigg|_{g\mu_B B_0} > 0
\end{multline}
is indeed positive as expected, because $\epsilon n_B(\epsilon)$ is a decreasing function. Therefore, in the magnon pumping process, higher-energy magnons pumped out of the left zeroth Landau level should result in more lower-energy magnons on the right zeroth Landau level with the assistance of electromagnetic energy injection. The total energy after pumping will then increase.

Before we leave this section, we calculate the anomalous spin and heat currents due to the magnon chiral anomaly. For the Weyl ferromagnet we considered, the spin and heat currents are given by the Landauer-B\"uttiker formalism \cite{meier2003},
\begin{multline}
J_{\text{spin}}^{\bm E, \bm B} = - \int_{g \mu_B B_R}^{+\infty}\hbar g_{\bm E}(\epsilon)n_B(\epsilon) v_R^{\bm E}(\epsilon) d\epsilon 
\\-  \int_{g \mu_B B_L}^{+\infty} \hbar g_{\bm E}(\epsilon)n_B(\epsilon) v_L^{\bm E}(\epsilon) d\epsilon,
\end{multline}
\begin{multline}
J_{\text{heat}}^{\bm E, \bm B} = \int_{g \mu_B B_R}^{+\infty} \epsilon g_{\bm E}(\epsilon)n_B(\epsilon) v_R^{\bm E}(\epsilon) d\epsilon 
\\+ \int_{g \mu_B B_L}^{+\infty} \epsilon g_{\bm E}(\epsilon)n_B(\epsilon) v_L^{\bm E}(\epsilon) d\epsilon,
\end{multline}
where the magnon drifting velocity in the zeroth Landau levels are $v_R^{\bm E}(\epsilon) = \frac{1}{\hbar}\frac{d\epsilon}{dq_z}|_R$ and $v_L^{\bm E}(\epsilon) = \frac{1}{\hbar}\frac{d\epsilon}{dq_z}|_L$. Further simplification gives the spin and heat currents as
\begin{equation} \label{spin_EB}
J_{\text{spin}}^{\bm E, \bm B}=-\hbar |v_z^\eta| (n_R^{\bm E, \bm B} - n_L^{\bm E, \bm B}) \approx \hbar\frac{g^2\mu_B^2}{2\pi^2 \hbar^2 c^2} n_B(g\mu_BB_0) B_5 \pazocal{E},
\end{equation}
\begin{equation} \label{heat_EB}
J_{\text{heat}}^{\bm E, \bm B} = |v_z^\eta| (U_R^{\bm E, \bm B} - U_L^{\bm E, \bm B}) \approx - \frac{g^3\mu_B^3}{2\pi^2 \hbar^2 c^2} n_B(g\mu_BB_0) B_0 B_5 \pazocal{E},
\end{equation}
where $v_R^{\bm E}(\epsilon) = -v_L^{\bm E}(\epsilon) = |v_z^\eta|$ is used. As discussed in Sec.~\ref{s3c}, the electric field $\bm E$ for magnons is dual to the vector potential $\bm A$ for electrons; the electric field gradient $\pazocal{E}$ thus plays the role of the magnetic field for electrons. $g\mu_BB_5$ measures the difference of the magnon population edges, and is therefore dual to the electron chiral chemical potential $\mu_5$. Consequently, the anomalous spin and heat currents of the magnon chiral anomaly is akin to the chiral magnetic current \cite{fukushima2008, liqiang2016} of the electron chiral anomaly. We will refer to Eqs.~\ref{spin_EB}, \ref{heat_EB} as the ``chiral electric effect''.

\subsection{Magnon chiral anomaly due to $\bm e^\eta$ and $\bm b^\eta$}
\label{s4b}
We now consider another possibility of implementing the magnon chiral anomaly. We use a chiral electric field $\bm e^\eta = \eta \bm e= (\frac{1}{2} \eta \varepsilon x, \frac{1}{2} \eta \varepsilon y, 0)$ induced by a static torsional strain, whose Landau levels are illustrated in Fig.~\ref{fig_6} with both zeroth Landau levels being right-moving. The twisted Weyl ferromagnet is aligned in the $z$ direction and in contact with magnon reservoirs, which will lift the magnon population edge to $-\bm \mu \cdot \bm B_0 = g\mu_BB_0$. As with the case of an electric field $\bm E$, in principle, the magnon population edge can be tuned into the gap of the first Landau levels (Fig.~\ref{fig_7}(d)) by a proper choice of $B_0$. Then a chiral magnetic field $\bm b^\eta = \eta \bm b = \eta b_z\hat z$, which is induced by a dynamic uniaxial strain, is overlaid, where $b_z$ has a nonzero gradient $\partial_zb_z=\beta$. Based on Eq.~\ref{EOM_bz}, the magnons are pumped through Landau levels according to the semiclassical equation of motion $q_z(t) = q_z(0)-\eta g\mu_B \int_0^t \beta t'/\hbar$. Thus magnons originally on the left zeroth Landau level are pumped into the right zeroth Landau level, giving rise to a magnon chiral anomaly.

Unlike the magnon pumping due to $\bm B$, magnons on different Weyl cones are pumped oppositely by the chiral magnetic field $\bm b^\eta$ such that the magnon population edge on left (right) zeroth Landau level is elevated (lowered) as illustrated in Fig.~\ref{fig_7}(e). The difference between magnon population edges can be characterized by a magnetic field bias $b_5 \ll B_0$ under which the left (right) Weyl cone experiences a magnetic field $b_L=B_0+b_5$ ($b_R=B_0-b_5$). According to the semiclassical equation of motion, we have
\begin{equation}
b_5 = -\frac{\hbar |v_z^\eta| \eta \int dq_z}{g\mu_B} = \int_0^t \beta |v_z^\eta|dt'.
\end{equation}
To derive the chiral anomaly due to $\bm e^\eta$ and $\bm b^\eta$, we need to know the magnon concentration variations on both zeroth Landau levels, which can be directly written down by referring to Eqs.~\ref{nR_EB}, \ref{nL_EB} as
\begin{multline} \label{nR_eb}
n_R^{\bm e, \bm b} = \frac{n_B(g\mu_B B_0)}{4\pi^2 l_e^2} \frac{g\mu_B b_5}{\hbar |v_z^\eta|} 
\\
-  \sum_{n=1} \frac{n_B^{(n)}(g\mu_B B_0)}{(n+1)!} \frac{(-g \mu_B b_5)^{n+1}}{4\pi^2 l_e^2 \hbar |v_z^\eta|}, 
\end{multline}
\begin{multline} \label{nL_eb}
n_L^{\bm e, \bm b} = - \frac{n_B(g\mu_B B_0)}{4\pi^2 l_e^2} \frac{g\mu_B b_5}{\hbar |v_z^\eta|} 
\\
- \sum_{n=1} \frac{n_B^{(n)}(g\mu_B B_0)}{(n+1)!} \frac{(g\mu_B b_5)^{n+1}}{ 4\pi^2 l_e^2 \hbar |v_z^\eta|}, 
\end{multline}
where the pseudo-electric length $l_e=(-\hbar c^2/g\mu_B\varepsilon )^{1/2}$. We assume $\sgn(g\varepsilon) = -1$ in order to be parallel to the assumption that $\sgn(g\pazocal{E})=-1$ (see Sec.~\ref{s3a}). In the low bias limit $g\mu_B b_5 \ll k_BT$, the chirality pumping rate is
\begin{multline} \label{QA_chiral_eb}
\frac{d\rho_5^{\bm e, \bm b}}{dt} = \chi_R \frac{dn_R^{\bm e, \bm b}}{dt} + \chi_L \frac{dn_L^{\bm e, \bm b}}{dt} 
\\
\approx - \frac{g^2\mu_B^2}{2\pi^2\hbar^2c^2}n_B(g \mu_B B_0) \varepsilon \beta.
\end{multline}
In the more general case, the magnon chiral anomaly equation can be written as
\begin{equation} \label{QA_eb}
\frac{d\rho_5^{\bm e, \bm b}}{dt} + \nabla \cdot j_5^{\bm e, \bm b} \approx \frac{n_B(g \mu_B B_0)}{2\pi^2\hbar^2c^2} \nabla(\bm \mu \cdot \bm b) \cdot [\nabla \times (\bm e \times \bm \mu)],
\end{equation}
which is parallel to Eq.~\ref{QA_EB}. Similarly, this anomaly equation only contains the leading order terms in Eqs.~\ref{nR_eb}, \ref{nL_eb}. More rigorously, $n_R^{\bm e, \bm b}+n_L^{\bm e, \bm b}>0$; otherwise the magnon pumping will become spontaneous. This can also be confirmed by checking the energy variations on both zeroth Landau levels. By referring to Eqs.~\ref{UR_EB}, \ref{UL_EB}, the estimated energy variations on the zeroth Landau levels are
\begin{multline} \label{UR_eb}
U_R^{\bm e, \bm b} \approx \frac{1}{4\pi^2 l_e^2 \hbar |v_z^\eta|} \bigg\{ \frac{1}{2} n_B(g\mu_B B_0) g^2\mu_B^2 [2B_0b_5 - b_5^2] 
\\
- n_B'(g\mu_B B_0) \frac{g\mu_B B_0 (-g\mu_B b_5)^2 }{2}  \bigg\},
\end{multline}
\begin{multline} \label{UL_eb}
U_L^{\bm e, \bm b} \approx \frac{1}{4\pi^2 l_e^2 \hbar |v_z^\eta|} \bigg\{ \frac{1}{2} n_B(g\mu_B B_0) g^2\mu_B^2 [ - 2B_0b_5-b_5^2] 
\\
- n_B'(g\mu_B B_0) \frac{g\mu_B B_0 (g\mu_B b_5)^2 }{2}\bigg\}.
\end{multline}
$U_R^{\bm e, \bm b}+U_L^{\bm e, \bm b}>0$, indicating energy injection due to the pseudo-EM fields during the pumping process.

\begin{figure*}[htb]
\includegraphics[width = 18cm]{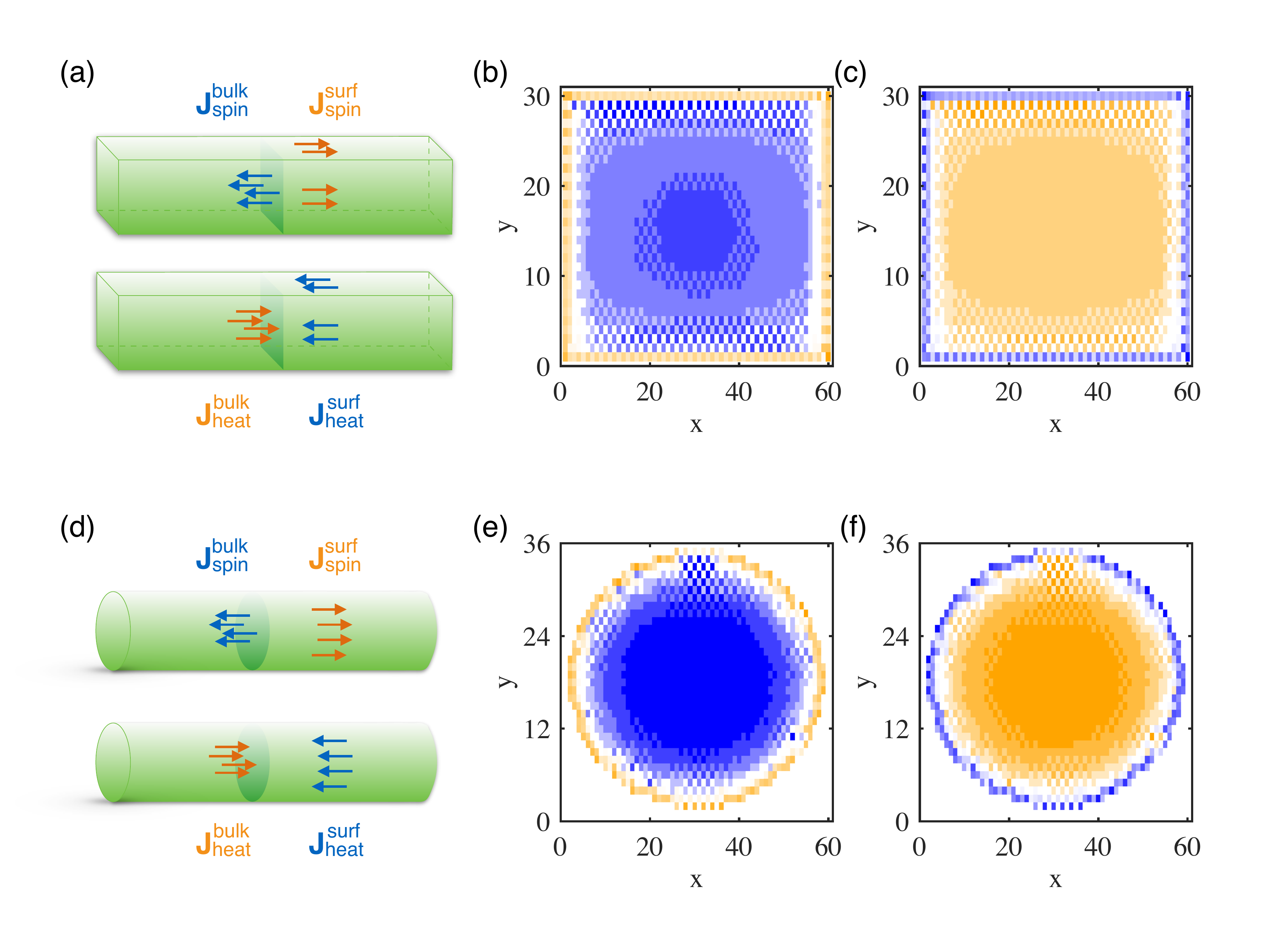}
\caption{Bulk-surface separation for the twisted Weyl ferromagnet nanowire. (a) Schematic plot of a Weyl ferromagnet nanowire with a rectangular cross section. The spin current propagates along the $-z$ direction in the bulk but along the $+z$ direction on the surface, while the heat current propagates along the $+z$ direction in the bulk but along the $-z$ direction on the surface. (b) Spatially resolved spin current on the cross section of the cuboid Weyl ferromagnet nanowire. (c) Spatially resolved heat current on the cross section of the cuboid Weyl ferromagnet nanowire. The directions of currents are color coded with blue (orange) representing $-z$ ($+z$). (d)-(f) Same as (a)-(c) but for a Weyl ferromagnet nanowire with an (almost) circular cross section. The total spin current on the rectangular (circular) cross section is $0.002DS$ ($0.0001DS$) while the total heat current on the rectangular (circular) cross section is $-0.0473D^2 S^2/\hbar$ ($-0.0018D^2 S^2/\hbar$). } \label{fig_8}
\end{figure*}

We now derive the anomalous spin and heat currents resulting from the chiral anomaly due to $\bm e^\eta$ and $\bm b^\eta$. First, it is important to note that before the chiral magnetic field $\bm b^\eta$ is switched on, there must be an equal number of right-moving magnons and left-moving magnons; otherwise the net spin/heat current will be nonzero in the absence of a driving force. As demonstrated in Sec.~\ref{s3b}, under the chiral electric field $\bm e^\eta$, the bulk only hosts right-moving magnons while the left-moving magnons are localized at the surface. Therefore, the bulk spin/heat current must be balanced by the surface spin/heat current when $\bm b^\eta = 0$. Explicitly, we have
\begin{multline}
\pazocal J_{\text{spin}}^{\text{bulk}} = -\int_{g\mu_B B_0 }^{+\infty} \hbar g_{\bm e}(\epsilon) n_B(\epsilon) v_R^{\bm e}(\epsilon) d\epsilon \\- \int_{g\mu_B B_0 }^{+\infty} \hbar g_{\bm e}(\epsilon) n_B(\epsilon) v_L^{\bm e}(\epsilon) d\epsilon = -\pazocal J_{\text{spin}}^{\text{surface}},
\end{multline}
\begin{multline}
\pazocal J_{\text{heat}}^{\text{bulk}} = \int_{g\mu_B B_0 }^{+\infty} \epsilon g_{\bm e}(\epsilon) n_B(\epsilon) v_R^{\bm e}(\epsilon) d\epsilon \\+ \int_{g\mu_B B_0 }^{+\infty} \epsilon g_{\bm e}(\epsilon) n_B(\epsilon) v_L^{\bm e}(\epsilon) d\epsilon = -\pazocal J_{\text{heat}}^{\text{surface}},
\end{multline}
where the density of states is $g_{\bm e}(\epsilon) = \frac{1}{2\pi l_e^2}\frac{1}{2\pi\hbar|v_z^\eta|}$ and the velocities are $v_R^{\bm e} = v_L^{\bm e} = |v_z^\eta|$. To verify our argument, we have numerically calculated the spatial distribution of magnon spin and heat currents on the cross section of the Weyl ferromagnet nanowire illustrated in the right panel of Fig.~\ref{fig_3}(a). As shown in Fig.~\ref{fig_8}(b), the spin current in the bulk of the rectangular cross section (Fig.~\ref{fig_3}(b)) propagates along the $-z$ direction while the spin current on the edges of the cross section propagates along the $+z$ direction. On the other hand, the bulk heat current of the rectangular cross section propagates along the $+z$ direction while the edge heat current propagates along the $-z$ direction, as illustrated in Fig.~\ref{fig_8}(c). We have carefully evaluated the total spin and heat currents through the rectangular cross section under various numerical settings, and find that both currents are vanishingly small when the gradient of $b_z$ vanishes, $\beta=0$. We further examine a Weyl ferromagnet nanowire with an almost circular cross section, whose bulk-surface separation for spin/heat transport is exhibited again in Fig.~\ref{fig_8} (e) and (f).

Then, we switch on the chiral pseudo-magnetic field $\bm b^\eta$, and the magnons begin to propagate along the Landau levels, giving rise to net anomalous spin and heat currents
\begin{multline}
J_{\text{spin}}^{\bm e, \bm b} = -\int_{g\mu_B b_R }^{+\infty} \hbar g_{\bm e}(\epsilon) n_B(\epsilon) v_R^{\bm e}(\epsilon) d\epsilon \\ - \int_{g\mu_B b_L }^{+\infty} \hbar g_{\bm e}(\epsilon) n_B(\epsilon) v_L^{\bm e}(\epsilon) d\epsilon - \pazocal J_{\text{spin}}^{\text{bulk}},  
\end{multline}
\begin{multline}
J_{\text{heat}}^{\bm e, \bm b} = \int_{g\mu_B b_R }^{+\infty} \epsilon g_{\bm e}(\epsilon) n_B(\epsilon) v_R^{\bm e}(\epsilon) d\epsilon \\ + \int_{g\mu_B b_L }^{+\infty} \epsilon g_{\bm e}(\epsilon) n_B(\epsilon) v_L^{\bm e}(\epsilon) d\epsilon - \pazocal J_{\text{heat}}^{\text{bulk}}. 
\end{multline}
Again, in the low bias limit $g\mu_B b_5 \ll k_BT$, we obtain the spin and heat currents to the lowest non-vanishing order
\begin{equation} \label{spin_eb}
J_{\text{spin}}^{\bm e, \bm b} = -\hbar |v_z^\eta| (n_R^{\bm e, \bm b} + n_L^{\bm e, \bm b}) 
\approx - \hbar \frac{g^3 \mu_B^3}{4\pi^2 \hbar^2 c^2} n'_B(g\mu_B B_0) b_5^2 \varepsilon,
\end{equation}
\begin{multline} \label{heat_eb}
J_{\text{heat}}^{\bm e, \bm b} = |v_z^\eta|(U_R^{\bm e, \bm b}+U_L^{\bm e, \bm b}) \\ \approx \frac{g^3 \mu_B^3}{4\pi^2\hbar^2 c^2}\Big[ n_B(g\mu_B B_0) + n'_B(g\mu_B B_0) g\mu_B B_0 \Big] b_5^2  \varepsilon. 
\end{multline}
Unlike the magnon chiral electric spin and heat currents Eqs.~\ref{spin_EB}, \ref{heat_EB}, which are proportional to the magnetic field bias $B_5$, the anomalous spin and heat currents due to $(\bm e^\eta, \bm b^\eta)$ are quadratic in the pseudo-magnetic field bias $b_5$, because magnons on both zeroth Landau levels have the identical velocity. These small but nonzero spin and heat currents reflect the non-conservation of magnon number, in contrast to the chiral anomaly of electrons with pure strain-induced pseudo-EM fields, where the anomalous current is exactly zero due to charge conservation.

\subsection{Magnon heat anomaly due to $\bm E$ and $\bm b^\eta$}
\label{s4c}
Besides the chiral anomalies discussed in Sec.~\ref{s4a} and Sec.~\ref{s4b}, Weyl magnons can exhibit another type of quantum anomaly in which the thermal energy in the bulk is not conserved. In this section, we will quantitatively characterize such a magnon ``heat anomaly" in the presence of an inhomogeneous electric field $\bm E$ and an inhomogeneous pseudo-magnetic field $\bm b^\eta$.

We again consider a Weyl ferromagnet nanowire subjected to an inhomogeneous electric field $\bm E = (\frac{1}{2} \pazocal{E} x, \frac{1}{2} \pazocal{E} y, 0)$. The two ends of the wire are attached to magnon reservoirs with a uniform magnetic field $\bm B_0 = B_0 \hat z$ such that the magnon population edge $g\mu_B B_0$ is located in the gap of the first Landau levels (Fig.~\ref{fig_7}(a)). Rather than using an inhomogeneous magnetic field $\bm B$ as is the case in Sec.~\ref{s4a}, we drive magnons with an inhomogeneous pseudo-magnetic field $\bm b^\eta = \eta \bm b = \eta b_z \hat z$, where $b_z$ has a nonzero gradient $\partial_z b_z = \beta$. As analyzed in Sec.~\ref{s4b}, the magnon motion is governed by the semiclassical equation of motion $q(t) = q(0) - \eta g\mu_B \int_0^t \beta dt' /\hbar$. Thus magnons are pumped into both zeroth Landau levels, indicating thermal energy non-conservation in the bulk, which is a clue of the magnon heat anomaly.

Due to the chiral nature of the pseudo-magnetic field $\bm b^\eta$, magnons on different zeroth Landau levels are oppositely pumped such that the magnon population edge on both zeroth Landau levels are lowered as illustrated in Fig.~\ref{fig_7}(c). The variation of the magnon population edge is denoted by $\delta_b \ll B_0$, and correspondingly both Weyl cones experience a magnetic field $b_L' = b_R' = B_0 - \delta_b$. According to the semiclassical equation of motion, we have
\begin{equation}
\delta_b=-\frac{\hbar |v_z^\eta| \eta \int dq_z}{g\mu_B} = \int_0^t \beta |v_z^\eta|dt'.
\end{equation}
By comparing to Eqs.~\ref{nR_EB}, \ref{nL_EB}, we can directly write down the magnon concentration variation on both zeroth Landau levels as 
\begin{equation} \label{nRnL_Eb}
n_R^{\bm E, \bm b} = n_L^{\bm E, \bm b} \approx   \frac{n_B(g\mu_BB_0)}{4\pi^2 l_E^2} \frac{g\mu_B\delta_b}{\hbar |v_z^\eta|},
\end{equation}
where we make the approximation $g\mu_B\delta_b \ll k_BT$. Because the magnon population edge is shifted identically on both chiral Landau levels, there is no net chirality transport between two Weyl cones. Nevertheless, the fact $n_R^{\bm E, \bm b}+n_L^{\bm E, \bm b}>0$ \cite{note2} indicates that there are more magnons in the bulk. Therefore, the thermal energy in the bulk increases. By comparing to Eqs.~\ref{UR_EB}, \ref{UL_EB}, we immediately obtain the heat injection into the two zeroth Landau levels in the limit $g\mu_B\delta_b \ll k_BT$ as
\begin{equation} 
U_R^{\bm E, \bm b}=U_L^{\bm E, \bm b}\approx \frac{n_B(g\mu_B B_0)}{4\pi^2 l_E^2}   g \mu_B B_0  \frac{g \mu_B \delta_b}{\hbar |v_z^\eta|}.
\end{equation}
Thus the bulk heat injection rate can be written down as
\begin{equation} \label{QA_heat_Eb}
\frac{d\rho_{\text{heat}}^{\bm E, \bm b}}{dt} =  \frac{dU_R^{\bm E, \bm b}}{dt} + \frac{dU_L^{\bm E, \bm b}}{dt}
\approx - \frac{g^3 \mu_B^3 n_B(g\mu_B B_0)}{2\pi^2 \hbar^2 c^2} B_0 \pazocal{E} \beta.
\end{equation}
More generally, the bulk heat injection rate reads
\begin{equation} \label{QA_Eb}
\frac{d\rho_{\text{heat}}^{\bm E, \bm b}}{dt}+\nabla \cdot \bm j_{\text{heat}}^{\bm E, \bm b} \approx \frac{g \mu_B B_0 n_B(g\mu_B B_0)}{2\pi^2 \hbar^2 c^2} \nabla (\bm \mu \cdot \bm b) \cdot [\nabla \times (\bm E \times \bm \mu)].
\end{equation}
This magnon heat anomaly equation is analogous to the magnon chiral anomaly equation (Eqs.~\ref{QA_EB}, \ref{QA_eb}). It is a heat continuity equation with a source indicating that the bulk thermal energy is not conserved. Unfortunately, this heat anomaly does not have measurable anomalous currents 
\begin{equation} \label{spin_Eb}
J_{\text{spin}}^{\bm E, \bm b} = 0,
\end{equation}
\begin{equation} \label{heat_Eb}
J_{\text{heat}}^{\bm E, \bm b} = 0.
\end{equation}
Because left-moving and right-moving magnons are always created/annihilated in pairs, as illustrated in Fig.~\ref{fig_7}(c). 

We note that though the bulk thermal energy is not conserved, the total energy of a closed system must be conserved. Since Eq.~\ref{QA_Eb} only characterizes the variation rate of the bulk thermal energy, we need to consider how the surface thermal energy is altered by the pseudo-magnetic field. As illustrated in Fig.~\ref{fig_4}(a), the bulk zeroth Landau levels are connected by a set of surface states. The magnons residing in these states can enter the bulk such that the thermal energy is transferred from the surface to the bulk. During this process, the external pseudo-magnetic field also does work on magnons; otherwise the heat pumping from surface to bulk would be spontaneous. The thermal energy from the surface and the mechanical energy from the pseudo-magnetic field constitute the heat injection into the bulk of the Weyl ferromagnet nanowire. In particular, if the pseudo-magnetic field is induced by the dynamic lattice deformation resulting from applying an ultrasonic sound wave, the energy loss during sound propagation in the Weyl ferromagnet will lead to sound attenuation, which, in principle, should be experimentally measurable.

\subsection{Magnon heat anomaly due to $\bm e^\eta$ and $\bm B$}
\label{s4d}
We now consider another possibility of implementing the magnon heat anomaly. We apply a strain-induced chiral pseudo-electric field $\bm e^\eta = \eta \bm e = \eta (\frac{1}{2} \varepsilon x, \frac{1}{2} \varepsilon y, 0)$ to a Weyl ferromagnet nanowire aligned in $z$ direction. The two ends of the wire are attached to magnon reservoirs subjected to a magnetic field $\bm B_0 = B_0\hat z$ with the magnon population edges lying in the gap of the first Landau levels
(Fig.~\ref{fig_7}(d)). We drive magnons with an inhomogeneous magnetic field $\bm B = B_z \hat z$, where $B_z$ has a nonzero gradient $\partial_z B_z = \pazocal{B}$. According to Sec.~\ref{s4a}, the magnon motion is governed by the semiclassical equation of motion $q_z(t) = q_z(0) - g\mu_B \int_0^t \pazocal{B} dt'/\hbar$. Consequently, magnons are pumped into both zeroth Landau levels and the thermal energy in the bulk increases, indicating a magnon heat anomaly.

The magnetic field $\bm B$ pumps magnons on different Weyl cones identically. However, due to the chiral nature of pseudo-electric field $\bm e^\eta$, the magnon population edge on both zeroth Landau levels are lowered as illustrated in Fig.~\ref{fig_7}(f). The variation of the magnon population edge is denoted as $\delta_B \ll B_0$ and both Weyl cones experience a magnetic field $B_L' = B_R' = B_0 - \delta_B$. According to the semiclassical equation of motion, we have
\begin{equation}
\delta_B = -\frac{\hbar |v_z^\eta| \int dq_z}{g\mu_B} = \int_0^t \pazocal{B} |v_z^\eta| dt'.
\end{equation}
By comparing to Eqs.~\ref{nR_EB}, \ref{nL_EB}, the magnon concentration variation on both zeroth Landau levels can be calculated as
\begin{equation} \label{nRnL_eB}
n_R^{\bm e, \bm B} = n_L^{\bm e, \bm B} \approx \frac{n_B(g\mu_B B_0)}{4\pi^2 l_e^2} \frac{g \mu_B \delta_B}{\hbar |v_z^\eta|},
\end{equation}
in the limit $g\mu_B\delta_B \ll k_BT$. Because magnons on both zeroth Landau levels are always created in pairs, there is no net chirality transport between the two Weyl points, but the larger total number of magnons on the zeroth Landau levels indicates a thermal energy injection into the bulk of the Weyl ferromagnet. By comparing to Eqs.~\ref{UR_EB}, \ref{UL_EB}, we can directly write down the heat injection into the two zeroth Landau levels in the limit $g\mu_B\delta_B \ll k_BT$ as
\begin{equation}
U_R^{\bm e, \bm B}=U_L^{\bm e, \bm B}\approx \frac{n_B(g\mu_B B_0)}{4\pi^2 l_e^2}   g \mu_B B_0  \frac{g \mu_B \delta_B}{\hbar |v_z^\eta|}.
\end{equation}
Thus the rate of heat injection into the bulk of the Weyl ferromagnet nanowire reads
\begin{equation} \label{QA_heat_eB}
\frac{d\rho_\text{heat}^{\bm e, \bm B}}{dt}=\frac{dU_R^{\bm e, \bm B}}{dt} + \frac{dU_L^{\bm e, \bm B}}{dt} \approx - \frac{g^3 \mu_B^3 n_B(g\mu_B B_0)}{2\pi^2 \hbar^2 c^2} B_0 \varepsilon \pazocal{B}.
\end{equation}
More generally, the bulk heat injection rate reads
\begin{equation} \label{QA_eB}
\frac{d\rho_\text{heat}^{\bm e, \bm B}}{dt} + \nabla \cdot \bm j_\text{heat}^{\bm e, \bm B} \approx \frac{g \mu_B B_0 n_B(g\mu_B B_0)}{2\pi^2 \hbar^2 c^2} \nabla(\bm \mu \cdot \bm B) \cdot [\nabla \times (\bm e \times \bm \mu)].
\end{equation}
which is similar to the heat anomaly equation (Eq.~\ref{QA_Eb}), showing non-conservation of the bulk thermal energy. Again, when we take into account the contribution of the surface states and the applied pseudo-electric field, the heat anomaly will be removed and the total energy is conserved. In contrast to the heat anomaly due to $\bm E$ and $\bm b^\eta$ in Sec.~\ref{s4c}, the heat anomaly due to $\bm e^\eta$ and $\bm B$ does result in anomalous spin and heat currents
\begin{equation} \label{spin_eB}
J_{\text{spin}}^{\bm e, \bm B}= -\hbar |v_z^\eta| (n_R^{\bm e, \bm B} + n_L^{\bm e, \bm B}) 
\approx \hbar \frac{g^2\mu_B^2}{2\pi^2\hbar^2c^2} n_B(g\mu_BB_0) \delta_B\varepsilon,
\end{equation}
\begin{equation} \label{heat_eB}
J_{\text{heat}}^{\bm e, \bm B}= |v_z^\eta| (U_R^{\bm e, \bm B} + U_L^{\bm e, \bm B})
\approx -\frac{g^3\mu_B^3}{2\pi^2\hbar^2c^2} n_B(g\mu_BB_0) B_0 \delta_B\varepsilon,
\end{equation}
both of which should be experimentally measurable. Unlike the magnon ``chiral electric effect" (Eqs.~\ref{spin_EB}, \ref{heat_EB}, \ref{spin_eb}, \ref{heat_eb}) whose spin and heat currents result from either EM fields or pseudo EM fields, the anomalous currents Eqs.~\ref{spin_eB}, \ref{heat_eB} result from their combination. Since the required pseudo-electric field $\bm e^\eta$ is induced by torsional strain, we refer to Eqs.~\ref{spin_eB}, \ref{heat_eB} as the magnon ``chiral torsional effect''.

\section{Field dependence of anomalous spin and heat currents}
\label{s5}
In Sec.~\ref{s4}, we listed the magnon quantum anomaly equations (Eqs.~\ref{QA_EB}, \ref{QA_eb}, \ref{QA_Eb}, \ref{QA_eB}) and the associated anomalous spin (Eqs.~\ref{spin_EB}, \ref{spin_eb}, \ref{spin_Eb}, \ref{spin_eB}) and heat (Eqs.~\ref{heat_EB}, \ref{heat_eb}, \ref{heat_Eb}, \ref{heat_eB}) currents. The anomaly equations have explicit EM/pesudo-EM field dependence while the anomalous currents do not, because the explicit field dependence of magnetic field bias ($B_5$/$b_5$) and magnetic field variation ($\delta_b$/$\delta_B$) is unknown. In this section, we will derive how these quantities depend on EM/pseudo-EM fields, and eventually give the full field dependence for the anomalous spin and heat currents in both the semiclassical limit and the quantum limit.

We take the magnon chiral anomaly due to $\bm E$ and $\bm B$ as an example and determine how $B_5$ depends on these fields. In the semiclassical limit, the magnetic field bias between the two zeroth Landau levels dominates over the magnon Landau spacing $ g\mu_B B_5 \gg \hbar \sqrt{ 2 | g\mu_B \pazocal{E} v_x^\eta v_y^\eta / \hbar c^2 |}$. Therefore, the chirality transported between the Weyl cones is
\begin{multline} \label{rho5_1}
\rho_5^{\bm E, \bm B}= \bigg\{ \chi_R \int_{g \mu_B B_R}^{+\infty} + \chi_L \int_{g \mu_B B_L}^{+\infty} \bigg\} D_s(\epsilon) n_B(\epsilon) d\epsilon
\\
\approx  \frac{g^2 \mu_B^2 B_0^2}{2\pi^2\hbar^3|v_x^\eta v_y^\eta v_z^\eta|} n_B(g \mu_B B_0) \cdot 2g\mu_B B_5,
\end{multline}
where the magnon density of states can be estimated using the dispersion $\varepsilon_{\bm k} = \hbar \sqrt{(v_x^\eta k_x)^2+(v_y^\eta k_y)^2+(v_z^\eta k_z)^2}$ as
$$
D_s(\epsilon) = \frac{1}{V} \sum_{\bm k} \delta(\epsilon - \varepsilon_{\bm k}) = \frac{\epsilon^2}{2\pi^2\hbar^3|v_x^\eta v_y^\eta v_z^\eta|}.
$$ 
In Sec.~\ref{s3a}, we have obtained the chirality pumping rate (Eq.~\ref{QA_chiral_EB}) in the absence of scattering of magnons between the two zeroth Landau levels. However, in a realistic magnet, the chirality mixing scattering mechanism always exists; otherwise the chirality transported between Weyl cones goes to infinity. Due to such scattering, the chirality pumping rate is changed to
\begin{equation} 
\frac{d \rho_5^{\bm E, \bm B}}{dt} = - \frac{g^2\mu_B^2}{2\pi^2\hbar^2c^2}n_B(g \mu_B B_0) \pazocal{EB} - \frac{\rho_5^{\bm E, \bm B}}{\tau^{\bm E,\bm B}},
\end{equation}
where $\tau^{\bm E,\bm B}$ is the mean free time of magnons due to chirality mixing scattering. The solution for $\rho_5^{\bm E, \bm B}$ at sufficiently long times $t \gg \tau^{\bm E,\bm B}$ is then
\begin{equation} \label{rho5_2}
\rho_5^{\bm E,\bm B} = - \frac{g^2\mu_B^2}{2\pi^2\hbar^2c^2}n_B(g \mu_B B_0) \pazocal{EB} \tau^{\bm E, \bm B}.
\end{equation}
By referring to Eq.~\ref{rho5_1}, we obtain the field dependence of the effective magnetic field bias as
\begin{equation}
B_5 = -\frac{\hbar |v_x^\eta v_y^\eta v_z^\eta|}{ 2g\mu_B c^2 B_0^2} \pazocal{EB} \tau^{\bm E, \bm B} \propto \pazocal{EB}. 
\end{equation}
where we have assumed that the chirality mixing scattering is characterized by a constant magnon mean free time $\tau^{\bm E,\bm B}$. In that case, the field dependence of the anomalous spin/heat current can be obtained as $J_{\text{spin/heat}}^{\bm E, \bm B} \propto \pazocal{E}^2 \pazocal{B}$, which is dual to the electron chiral magnetic current since magnon EM fields are dual to electron EM potentials. 

In the quantum limit, the Landau level spacing becomes comparable to the magnetic field bias $ g\mu_B B_5 \lesssim \hbar \sqrt{ 2 | g\mu_B \pazocal{E} v_x^\eta v_y^\eta / \hbar c^2 |}$, and the magnon density of states is proportional to the electric field gradient ($D_q(\epsilon) \sim 1/l_E^2 \propto \pazocal{E}$). The chirality transported between Weyl cones is then $\rho_5^{\bm E,\bm B} \propto \pazocal{E} B_5$, distinct from the semiclassical result Eq.~\ref{rho5_1}. By comparing to Eq.~\ref{rho5_2}, we obtain the field dependence of the magnetic field bias $B_5 \propto \pazocal{B}$, provided that the magnon mean free time is constant. As a result, the field dependence of the anomalous spin/heat current in the quantum limit can be immediately obtained as $J_{\text{spin/heat}}^{\bm E, \bm B} \propto \pazocal{E} \pazocal{B}$.

By replicating the analysis above, we can determine the field dependence of other magnetic field bias/variations as $b_5 \propto \varepsilon \beta$, $\delta_b \propto \pazocal{E} \beta$, and $\delta_B \propto \varepsilon \pazocal{B}$ in the semiclassical limit and $b_5 \propto \beta$, $\delta_b \propto \beta$, and $\delta_B \propto \pazocal{B}$ in the quantum limit. The resulting field dependence of the corresponding spin/heat current is summarized in Table.~\ref{tab1}. The magnon chiral electric spin/heat current (Eqs.~\ref{spin_EB}, \ref{heat_EB}) is dual to the chiral magnetic current \cite{fukushima2008, liqiang2016}, and the magnon chiral torsional spin/heat current (Eqs.~\ref{spin_eB}, \ref{heat_eB}) is dual to the chiral torsional current given in Ref.~\cite{pikulin2016}. However, it is worth noting that the magnon anomalous spin/heat current (Eqs.~\ref{spin_eb}, \ref{heat_eb}) due to $\bm e$ and $\bm b$ shows unprecedented field dependence without any anomalous electric current counterpart.

\begin{table}[htb] 
\caption{Summary of field (gradient) dependence of anomalous spin and heat currents in magnon quantum anomalies.} \label{tab1}
\begin{tabular}{C{2cm}C{3.1cm}C{3.1cm}}
\hline\hline \\[-1em]
\textbf{Field} & \textbf{Semiclassical limit} & \textbf{Quantum limit}\\ \\[-1em]
\hline \\[-0.5em]
   
    $\bm E, \bm B$ & 
    $J_{\text{spin/heat}}^{\bm E, \bm B} \propto \pazocal{E}^2 \pazocal{B}$ & 
    $J_{\text{spin/heat}}^{\bm E, \bm B} \propto \pazocal{E} \pazocal{B}$ \\  \\[-0.5em]  

    $\bm e, \bm b$ & 
    $J_{\text{spin/heat}}^{\bm e, \bm b} \propto \varepsilon^3 \beta^2$ & 
    $J_{\text{spin/heat}}^{\bm e, \bm b} \propto \varepsilon \beta^2$ \\ \\[-0.5em]
  
    $\bm E, \bm b$ & 
    $J_{\text{spin/heat}}^{\bm E, \bm b} =0$ &
    $J_{\text{spin/heat}}^{\bm E, \bm b} =0$ \\ \\[-0.5em]

    $\bm e, \bm B$ & 
    $J_{\text{spin/heat}}^{\bm e, \bm B} \propto \varepsilon^2 \pazocal{B}$ &
   $J_{\text{spin/heat}}^{\bm e, \bm B} \propto \varepsilon \pazocal{B}$ \\  \\[-0.5em]
\hline\hline
\end{tabular}
\end{table}

\section{Experimental measurement of magnon quantum anomalies}
\label{s6}
In Sec.~\ref{s5}, we have obtained the field (gradient) dependence of anomalous spin and heat currents in magnon quantum anomalies as summarized in Tab.~\ref{tab1}. Unfortunately, the direct measurement of such anomalous spin and heat transport is experimentally non-trivial due to the lack of induced electric field from spin currents and the dissipation of thermal energy from heat currents. For this reason, an easily measurable signature quantity is required for the experimental detection of magnon quantum anomalies. 

We propose that the force on the magnon current-carrying Weyl ferromagnet nanowire exerted by an external inhomogeneous electric or pseudo-electric field could be such a signature quantity. Intuitively, this force on a magnon current can be understood as an analog of Amp\`ere force, which emerges when applying an external magnetic field to an electric current. In the following, we will derive this force quantitatively for each magnon quantum anomaly.

\subsection{Experimental signature of magnon chiral anomaly due to  $\bm E$ and $\bm B$}
\label{s6a}
In Sec.~\ref{s3a}, we have demonstrated that magnon bands are Landau-quantized by an external inhomogeneous electric field $\bm E = (\frac{1}{2} \pazocal E x, \frac{1}{2} \pazocal E y, 0)$. We now apply an additional electric field $\bm E' = (0, \pazocal E'z, 0)$ with $\pazocal E' \ll \pazocal E$ to the Weyl ferromagnet nanowire. This weak additional electric field will not lead to further Landau quantization of magnon bands, but according to the magnon equation of motion (Eq.~\ref{EOM_BE}), it gives rise to a ``magnon Lorentz force"
\begin{equation} \label{EOM_Ep}
\hbar \frac{d\bm k}{dt} = \frac{1}{c^2} \bm v \times [\nabla \times (\bm \mu \times \bm E')],
\end{equation}
which results from the Aharonov-Casher effect \cite{aharonov1984}. For the magnon population illustrated in Fig.~\ref{fig_7}(a), the net force contributed by magnons from the two zeroth Landau levels is zero, because for each magnon drifting at velocity $|v_z^\eta|$, there is always another magnon drifting at velocity $-|v_z^\eta|$. 

However, when magnons are pumped by an inhomogeneous magnetic field $\bm B = B_z \hat z$ in the $z$ direction, the net force on the whole nanowire is no longer zero, thus serving as an experimental signature for the magnon chiral anomaly due to $\bm E$ and $\bm B$. This is because a magnon imbalance develops on the zeroth Landau levels as illustrated in Fig.~\ref{fig_7}(b) such that there are more right-moving magnons than left-moving ones. In addition to causing the magnon chiral anomaly (Eq.~\ref{QA_EB}) and the associated anomalous transport (Eqs.~\ref{spin_EB}, \ref{heat_EB}), this magnon imbalance also produces a force
\begin{multline} \label{F_EB}
\bm F^{\bm E, \bm B} =  \sum_{LL_0} \frac{1}{c^2} \bm v \times [\nabla \times (\bm \mu \times \bm E')]
\\
=-V(n_R^{\bm E, \bm B} - n_L^{\bm E, \bm B})|v_z^\eta| \frac{g\mu_B\pazocal E'}{c^2} \hat x,
 \end{multline}
where the summation goes over the zeroth Landau levels ($LL_0$) and $V$ refers to the volume of the nanowire. To estimate the magnitude of this force, we first assume the additional electric field gradient $\pazocal E' = 0.1 \pazocal E$ such that $\pazocal E'$ has little effect on the magnon band structure in Fig.~\ref{fig_4}. We then take the magnon drifting velocity $|v_z^\eta| \sim 10^2$m/s of the same order as that of yttrium iron garnet (YIG) \cite{xie2017}. For a typical nanowire with cross section radius $\sim 100$nm and length $\sim 100 \mu$m, the volume can be estimated as $V \sim 10^{-18}$m$^3$. To estimate the magnon concentration imbalance, we make use of Eqs.~\ref{nR_EB}, \ref{nL_EB} and get
\begin{equation}
n_R^{\bm E, \bm B} - n_L^{\bm E, \bm B} \approx\frac{n_B(g\mu_BB_0)}{2\pi^2l_E^2}\frac{g\mu_B\pazocal B}{\hbar} \tau^{\bm E, \bm B}.
\end{equation}
For the electric field gradient used in Fig.~\ref{fig_4},  we obtain $1/2\pi^2l_E^2 \sim 10^{14}$m$^{-2}$. We further estimate the magnon mean free time $\tau^{\bm E, \bm B} \sim 10^{-6}$s, which is of the same order as that of YIG \cite{douglass1963}. Lastly, an inhomogeneous magnetic field with gradient $\pazocal B \sim 10$T/m should be experimentally available. These lead to a magnon concentration imbalance $n_R^{\bm E, \bm B} - n_L^{\bm E, \bm B} \sim 10^{20}$m$^{-3}$ and a force $F^{\bm E, \bm B} \sim 10^{-15}$N. By means of atomic force microscopy (AFM), this small force can be sensed as a clue of magnon chiral anomaly due to $\bm E$ and $\bm B$. 

Before we leave this subsection, we briefly analyze the mechanical effects of magnons on the surface and the higher Landau levels. According to Fig.~\ref{fig_4}(a, b), the almost flat surface states indicate a vanishing magnon drifting velocity, thus a negligible force contribution is expected from the surface magnons. For the higher Landau levels, the magnon population alters little during the whole pumping process because all states in these bands are occupied. The force contribution is thus ideally zero because the numbers of right-moving and left-moving magnons are equal. In conclusion, the net force on the nanowire is mostly contributed by the magnon imbalance on the zeroth Landau levels.

\subsection{Experimental signature of magnon chiral anomaly due to $\bm e^\eta$ and $\bm b^\eta$}
\label{s6b}
In Sec.~\ref{s3b}, we have demonstrated that magnon bands are Landau-quantized by an inhomogeneous pseudo-electric field $\bm e^\eta =  \eta (\frac{1}{2} \varepsilon x, \frac{1}{2} \varepsilon y, 0)$ induced by a static twist. Again, by applying an additional electric field $\bm E' = (0, \pazocal E'z, 0)$, a magnon will experience a Lorentz force given by Eq.~\ref{EOM_Ep}. However, for the magnon population illustrated in Fig.~\ref{fig_7}(d), the net force contributed by magnons on the two zeroth Landau levels is nonzero because these two bands are now co-propagating. 

Nevertheless, when magnons are pumped by a pseudo-magnetic field in the $z$ direction $\bm b^\eta = \eta b_z \hat z$, a magnon imbalance develops as illustrated in Fig.~\ref{fig_7}(e), with slightly more right-moving magnons on the zeroth Landau levels. The force acting on the nanowire then increases by
\begin{equation}
\delta \bm F^{\bm e, \bm b} = -V(n_R^{\bm e, \bm b} + n_L^{\bm e, \bm b})|v_z^\eta| \frac{g\mu_B\pazocal E'}{c^2}\hat x,
\end{equation}
where the magnon concentration variation can be estimated by Eqs.~\ref{nR_eb}, \ref{nL_eb} as
\begin{multline}
n_R^{\bm e, \bm b} + n_L^{\bm e, \bm b} \approx \frac{n_B(g\mu_BB_0)}{2\pi^2 l_e^2} \frac{g\mu_B\beta}{\hbar}\tau^{\bm e, \bm b} 
\\
\times \frac{e^{g\mu_BB_0/k_BT}}{e^{g\mu_BB_0/k_BT}-1} \frac{g\mu_B\beta |v_z^\eta| \tau^{\bm e, \bm b}}{2k_BT}.
\end{multline}
If we assume similar parameters, i.e., $l_e \sim l_E$, $\beta \sim \pazocal B$, and $\tau^{\bm e, \bm b} \sim \tau^{\bm E, \bm B}$, we can estimate the force increase as $\delta F^{\bm e, \bm b} \sim 10^{-21}$N at room temperature. Though this force increase reflects the magnon number non-conservation as well as the magnon anomalous transport (Eqs.~\ref{spin_eb}, \ref{heat_eb}), the difficulty of force sensing is greatly increased. 

To avoid this difficulty, we propose the force sensing experiment using an additional pseudo-electric field $\bm e'_\eta = \eta (0, \varepsilon'z, 0)$, which can be generated by a circular bend deformation (Appendix~\ref{a4}). For the magnon population shown in Fig.~\ref{fig_7}(d), the total Lorentz force on magnons on the zeroth Landau levels is restored to zero due to the chiral nature of the additional pseudo-electric field. However, as the pseudo-magnetic field $\bm b^\eta$ is switched on, the magnon imbalance illustrated in Fig.~\ref{fig_7}(e) will produce a nonzero force $F^{\bm e, \bm b}$ acting on the whole nanowire. Explicitly, the force reads
\begin{equation}
\bm F^{\bm e, \bm b} = -V(n_R^{\bm e, \bm b} - n_L^{\bm e, \bm b})|v_z^\eta| \frac{g\mu_B\varepsilon'}{c^2} \hat x,
\end{equation} 
where the magnon concentration imbalance is given by Eqs.~\ref{nR_eb}, \ref{nL_eb} as
\begin{equation}
n_R^{\bm e, \bm b} - n_L^{\bm e, \bm b} \approx \frac{n_B(g\mu_BB_0)}{2\pi^2l_e^2}\frac{g\mu_B\beta}{\hbar}\tau^{\bm e, \bm b}.
\end{equation}
Using the parameters above, we obtain the magnon imbalance $n_R^{\bm e, \bm b} - n_L^{\bm e, \bm b} \sim 10^{20}$m$^{-3}$ and the force exerted on the nanowire is then $F^{\bm e, \bm b} \sim 10^{-15}$N, which can be measured by AFM. It is worth noting that the surface magnons, despite possessing a finite drifting velocity, do not have an appreciable force contribution, because the additional pseudo-electric field $\bm e'_\eta$ only lives in the vicinity of Weyl cones deep in the bulk and thus has no effect on the surface. Unlike $\delta F^{\bm e, \bm b}$ which reflects the anomalous magnon transport (Eqs.~\ref{spin_eb}, \ref{heat_eb}), the force $F^{\bm e, \bm b}$ indicates the magnon population imbalance on the zeroth Landau levels and is thus a signature of the magnon chiral anomaly due to $\bm e^\eta$ and $\bm b^\eta$.

\subsection{Experimental signature of magnon heat anomaly due to $\bm E$ and $\bm b^\eta$}
\label{s6c}
For the magnon heat anomaly due to $\bm E$ and $\bm b^\eta$, the Landau quantization is still provided by the inhomogeneous electric field, but the pumping field is a chiral pseudo-magnetic field due to a dynamic uniaxial strain, leading to an equal number of left-moving and right-moving magnons injected into the bulk as illustrated in Fig.~\ref{fig_7}(c). This renders $\bm E' = (0, \pazocal E' z, 0)$ in Sec.~\ref{s6a} useless because the force contributed by the magnons on the zeroth Landau levels is always zero. However, an additional pseudo-electric field $\bm e'_\eta = \eta(0, \varepsilon' z, 0)$ produces a measurable mechanical effect.

Due to the chiral nature of $\bm e'_\eta$, the magnons on the zeroth Landau levels (Fig.~\ref{fig_7}(a)) contribute constructively to the force exerted on the nanowire. When the magnons are pumped by the pseudo-magnetic field, more magnons are driven into the zeroth Landau levels, leading to an increase in this force
\begin{equation}
\delta \bm F^{\bm E, \bm b} = -V (n_R^{\bm E, \bm b} + n_L^{\bm E, \bm b}) |v_z^\eta| \frac{g\mu_B\varepsilon'}{c^2} \hat x,
\end{equation}
where the magnon concentration variation is given by Eq.~\ref{nRnL_Eb} as
\begin{equation}
n_R^{\bm E, \bm b}+n_L^{\bm E, \bm b} \approx \frac{n_B(g\mu_BB_0)}{2\pi^2l_E^2} \frac{g\mu_B \beta}{\hbar} \tau^{\bm E, \bm b}. 
\end{equation}
Assuming $\tau^{\bm E, \bm b}\sim 10^{-6}$s leads to an AFM-measurable force increase $\delta F^{\bm E, \bm b} \sim 10^{-15}$N. As analyzed in Sec.~\ref{s6a}, this force is the actual force experienced by the nanowire because there is no force contribution from the magnons on the surface or the higher Landau levels. The force increase $\delta F^{\bm E, \bm b}$ is of great experimental importance, because there is no magnon anomalous transport (Eqs.~\ref{spin_Eb}, \ref{heat_Eb}) in the magnon quantum anomaly due to $\bm E$ and $\bm b^\eta$.

\subsection{Experimental signature of magnon heat anomaly due to $\bm e^\eta$ and $\bm B$}
\label{s6d}
For the magnon heat anomaly due to $\bm e^\eta$ and $\bm B$, the Landau quantization is provided by the inhomogeneous pseudo-electric field resulting from a static twist, but the pumping field is an ordinary magnetic field, leading to an equal number of magnons injected into each zeroth Landau level as illustrated in Fig.~\ref{fig_7}(f). Consequently, the magnons on the zeroth Landau levels contribute zero total force in the presence of the chiral pseudo-electric field $\bm e'_\eta = \eta(0, \varepsilon'z, 0)$. We thus resort to an additional ordinary electric field $\bm E' = (0, \pazocal E'z, 0)$ as is used in Sec.~\ref{s6a}.

Because the two zeroth Landau levels are co-propagating under the pseudo-electric field $\bm e^\eta$, which results from a static twist, the magnons on these two zeroth Landau levels contribute constructively to the force exerted on the nanowire in the presence of the additional electric field $\bm E'$. When more magnons are pumped by $\bm B$ into the zeroth Landau levels, the force increases by
\begin{equation}
\delta \bm F^{\bm e, \bm B} = -V (n_R^{\bm e, \bm B} + n_L^{\bm e, \bm B}) |v_z^\eta| \frac{g\mu_B\pazocal E'}{c^2} \hat x,
\end{equation}
where the magnon concentration variation is given by Eq.~\ref{nRnL_eB} as
\begin{equation}
n_R^{\bm e, \bm B}+n_L^{\bm e, \bm B} \approx \frac{n_B(g\mu_BB_0)}{2\pi^2l_e^2} \frac{g\mu_B \pazocal B}{\hbar} \tau^{\bm e, \bm B}.
\end{equation}
Assuming $\tau^{\bm e, \bm B} \sim 10^{-6}$s leads to an AFM-measurable force increase $\delta F^{\bm e, \bm B} \sim 10^{-15}$N as a signature of both the magnon heat anomaly (Eq.~\ref{QA_eB}) and the anomalous transport (Eqs.~\ref{spin_eB}, \ref{heat_eB}). Unlike the case analyzed in Sec.~\ref{s6a} and Sec.~\ref{s6c} where surface magnons have a vanishing drifting velocity or the case analyzed in Sec.~\ref{s6b} where the effective surface pseudo-electric field is zero, the surface magnons in heat anomaly due to $\bm e^\eta$ and $\bm B$ give rise to non-trivial mechanical effects due to their non-vanishing drifting velocity and effective surface electric field $\bm E'$. However, because surface states and the zeroth Landau levels are counter-propagating and their magnon numbers always change in opposite directions, the force due to surface magnon concentration variation always adds constructively to the bulk force variation $\delta F^{\bm e, \bm B}$, leading to an even larger AFM-measurable mechanical effect.

\section{Conclusions}
\label{s7}
In this paper, we have derived the magnon quantum anomalies and the anomalous spin and heat currents in a Weyl ferromagnet under electromagnetic fields and strain-induced chiral pseudo-electromagnetic fields. We first analyze a multilayer model of a Weyl ferromagnet whose spin wave structure possesses two linearly dispersive Weyl cones located on the $k_z$ axis at the corners of the honeycomb lattice Brillouin zone, akin to the electronic structure of Weyl semimetals. We show that the two Weyl cones are connected by a set of surface states analogous to the ``Fermi arcs" in Weyl semimetals. These surface states can be understood as the combination of chiral edge states of each $k_z$-fixed 2D slice of the Weyl ferromagnet, which realizes a magnon Chern insulator whose Chern number is nontrivial for the momenta between the two Weyl points.

We then analyze how the Weyl ferromagnet reacts to EM fields and strain-induced chiral pseudo-EM fields. Under an inhomogeneous electric field $\bm E$, due to the Aharonov-Casher effect \cite{aharonov1984}, the magnons will be Landau-quantized. Similar Landau quantization can be obtained by a static twist (around $z$ axis) of the Weyl ferromagnet nanowire because an inhomogeneous pseudo-electric field $\bm e^\eta$ is induced. Such a chiral pseudo-electric field only lives in the vicinity of the magnon Weyl points and couples to them oppositely, leading to a pair of co-propagating zeroth Landau levels. The Landau-quantized magnons can be manipulated by applying an inhomogeneous magnetic field $\bm B$, which contributes a Zeeman energy whose gradient acts as a driving force. A similar pumping process is realized by applying a dynamic uniaxial strain to the Weyl ferromagnet, so that an inhomogeneous chiral pseudo-magnetic field $\bm b^\eta$ is induced. Again, this field strongly depends on the ``Diracness" of the spin wave structure and only couples to Weyl magnons. Due to its chiral nature, magnons are pumped oppositely on different Weyl cones.

Furthermore, we show that the four possible combinations of electric field ($\bm E$/$\bm e^\eta$) and magnetic field ($\bm B$/$\bm b^\eta$) give rise to magnon quantum anomalies and anomalous spin and heat currents. For $(\bm E, \bm B)$, magnons are pumped from one zeroth Landau level to the other, resulting in a chirality imbalance between the two Landau levels. Anomalous spin and heat currents arise and are proportional to this imbalance, resembling the chiral magnetic effect in Weyl semimetals. We thus call this anomaly-related transport the magnon ``chiral electric effect". For $(\bm e^\eta, \bm b^\eta)$, magnons are also injected into one zeroth Landau level and extracted out of the other, leading to a chirality imbalance as well. Remarkably, this magnon chiral anomaly due to pure pseudo-electromagnetic fields has weak but non-zero anomalous spin and heat currents, unlike the electron chiral magnetic current which must be zero in the presence of pseudo-electromagnetic fields. This is because magnons are quasiparticles immune to the particle conservation law. For $(\bm E, \bm b^\eta)$, magnons are pumped between the surface and the bulk, giving rise to a heat imbalance between the two. For this reason, we refer to such a phenomenon as the magnon ``heat anomaly" because the energy in the bulk itself is not conserved. Unfortunately, there are always an equal number of right-moving magnons and left-moving magnons; thus no net spin and heat currents exist. For $(\bm e^\eta, \bm B)$, magnons are also pumped between the surface and the bulk. Due to the chiral nature of $\bm e^\eta$, the bulk and the surface always have opposite velocities. Such a bulk-surface separation thus causes a magnon ``chiral torsional effect" that gives spin and heat currents proportional to the bulk-surface magnon imbalance.

Lastly, considering the difficulty of direct measurement of magnon anomalous spin and heat currents, we propose AFM-based force sensing experiments. For $(\bm E, \bm B)$, an additional electric field on the nanowire exerts a force $\bm F^{\bm E, \bm B}$ caused by the magnon imbalance on the zeroth Landau levels, reflecting the magnon chiral anomaly and the magnon chiral electric effect. For $(\bm e^\eta, \bm b^\eta)$, an additional pseudo-electric field on the nanowire gives rise to a force $\bm F^{\bm e, \bm b}$, also caused by the magnon imbalance on zeroth Landau levels, indicating the magnon chiral anomaly due to pseudo-EM fields. For $(\bm E, \bm b^\eta)$, we demonstrate that the additional pseudo-electric field causes a force increase $\delta \bm F^{\bm E, \bm b}$ on the nanowire, which is a signature of the magnon heat anomaly. This allows us to detect such an anomaly experimentally though the anomalous transport is lacking. For $(\bm e^\eta, \bm B)$, we elucidate that the additional electric field produces a force increase $\delta F^{\bm e, \bm B}$ on the nanowire, which is a clue of the magnon heat anomaly and the magnon chiral torsional effect.

To experimentally test the magnon quantum anomalies, we first require a Weyl ferromagnet, which may be artificially engineered by layering the honeycomb ferromagnet CrX$_3$ (X = F, Cl, Br, I) \cite{huang2017, pershoguba2018}. Weyl ferromagnets are also proposed to occur intrinsically in the pyrochlore oxide Lu$_2$V$_2$O$_7$ \cite{su2017b}. However, our theory constructed for multilayer ferromagnets cannot be directly transplanted to the pyrochlore lattice, where a twist around $[111]$ rather than $z$ axis may be the natural choice. The second requirement is that the candidate materials should be flexible to allow for sufficient twist in order to generate a strong pseudo-electric field to Landau-quantize the magnon bands. Unfortunately, the mechanical properties regarding the flexibility of candidate materials are lacking, and further experimental work is needed to verify whether or not layered honeycomb ferromagnets and Lu$_2$V$_2$O$_7$ are suitable materials.

There are several future directions that might be interesting to pursue based on the present work. The first is to test whether other types of spin lattice deformation can induce chiral pseudo-EM fields. In the context of Weyl semimetals, Ref.~\cite{sumiyoshi2016} shows that a screw dislocation can lead to the chiral torsional effect. It will be interesting to examine whether such deformation design will induce a pseudo-electric field in a Weyl magnet. Another direction is to study how other Weyl bosons react to strain-induced chiral gauge fields. To date, strain-induced Landau levels have been observed in photonic graphene \cite{rechtsman2013}. Since ``photonic Weyl semimetals" \cite{gao2016, chen2016, lin2016} have been proposed and realized, checking whether or not strain can induce chiral anomalies in these photonic systems will also be rewarding.

\begin{acknowledgments}
The authors are indebted to M. Franz, Y. Su, A. Furusaki, S. Fujimoto, and A. Chen for the insightful discussions. Z. Shi is supported by project A02 of the CRC-TR 183. We thank NSERC, CIfAR, and Max Planck - UBC Centre for Quantum Materials for support.
\end{acknowledgments}

\appendix
\section{Modulation of exchange integral under strain}
\label{a1}
In the context of electronic crystals, the effect of strain is to alter the position and the orientation of each electronic orbit. Microscopically, this modulation can be incorporated by a suitable substitution \cite{shapourian2015} of the overlap integral $t(\bm R' - \bm R)$, which only depends on the relative displacement of electronic orbits located at $\bm R'$ and $\bm R$. Similar to the overlap integral, the exchange integral $J(\bm R' - \bm R)$ depends on the relative displacement of spins located at $\bm R$ and $\bm R'$. Under strain, the spin originally located at $\bm R$ will be shifted to a new position $\bm R + \bm u(\bm R)$. By expanding to the linear order, the exchange integral is given by
\begin{multline}
J(\bm R' + \bm u(\bm R') - \bm R - \bm u(\bm R)) 
\\
\approx J(\bm R' - \bm R) + \nabla J \cdot (\bm u(\bm R')  - \bm u(\bm R))
\\
\approx J(\bm R' - \bm R) + \nabla J \cdot \big[ (\bm R' - \bm R) \cdot \nabla \bm u \big].
\end{multline}
We may estimate $\nabla J =-\alpha J\frac{\bm R' - \bm R}{(\bm R' - \bm R)^2} \approx -J\frac{\bm \delta}{\bm \delta^2}$, where we have taken the Gr\"uneisen parameter $\alpha \approx 1$ and denoted $\bm \delta = \bm R' - \bm R$. Then the substitution for the exchange integral is
\begin{equation} \label{sub_gen}
J(\bm \delta) \rightarrow J(\bm \delta) \Big(1-\frac{1}{\delta^2} \bm \delta \cdot \bm \delta \cdot  \nabla \bm u \Big),
\end{equation}
where $\bm \delta \cdot \bm \delta \cdot  \nabla \bm u = \delta_i \delta_j \partial_j u_i = \delta_i \delta_j u_{ij}$ and $u_{ij} = \frac{1}{2}(\partial_i u_j + \partial_j u_i)$ is the symmetrized strain tensor. For a twist deformation, the non-vanishing components of the strain tensor are $u_{31}$ that couples to $\delta_1$ and $\delta_3$, and $u_{32}$ that couples to $\delta_2$ and $\delta_3$. Therefore, only the exchange integrals whose arguments simultaneously have out-of-layer and in-layer components are affected. This refers to the six $J_2$'s as illustrated in Fig.~\ref{fig_5} in the main text. On the other hand, for a uniaxial deformation characterized by $u_{33}$ that couples to $\delta_3$, all exchange integrals whose arguments have inter-layer components will be changed. This refers to $J_A$, $J_B$, and the six $J_2$'s.

\section{Magnon bands in the presence of $\bm E$ and $\bm e^\eta$}
\label{a2}
In the main text, we have seen that the magnon bands of the Weyl ferromagnet can be Landau-quantized by either applying an inhomogeneous electric field $\bm E$ (Sec.~\ref{s3a}) or applying a twist which induces an inhomogeneous pseudo-electric field $\bm e^\eta$ (Sec.~\ref{s3b}). The former acts on the whole Weyl ferromagnet, while the latter only couples to Weyl magnons and is greatly suppressed at higher energies where the ``Diracness" of magnon bands is no longer preserved. Moreover, the pseudo-electric field is chiral and oppositely couples to different Weyl cones, resulting in two identically dispersing zeroth Landau levels, as illustrated in Fig.~\ref{fig_6}(c) in the main text. 

\begin{figure*}[htb]
\includegraphics[width = 16cm]{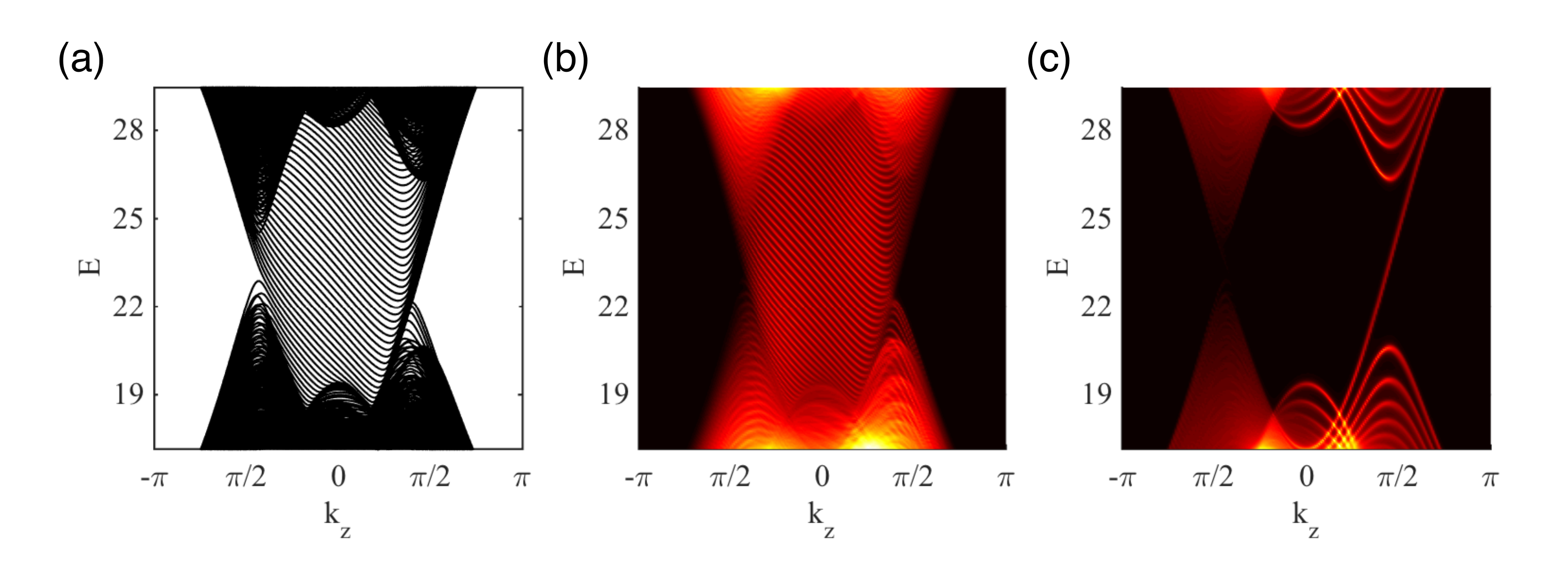}
\caption{Magnon dispersion for a twisted Weyl ferromagnet nanowire in the presence of an inhomogeneous electric field. For all panels, we take $\frac{g\mu_B \pazocal{E} a^2}{ec^2}=\frac{g\mu_B \varepsilon a^2}{ec^2}=-0.0124\Phi_0$. (a) Magnon band structure. Due to the chiral nature of the strain-induced pseudo-electric field, the effective electric field at the left Weyl cone vanishes while the effective field at the right Weyl cone is doubled. Therefore, the left Weyl cone is not Landau-quantized but the right Weyl cone exhibits Dirac-Landau levels. (b) Surface spectral function, which shows a set of left-moving surface states connecting the left Weyl cone and the right zeroth Landau level. (c) Bulk spectral function, which clearly unveils the linear band touching at the left Weyl cone, and Dirac-Landau levels at the right Weyl cone. Compared to Fig.~\ref{fig_4}(c) and Fig.~\ref{fig_6}(c), the Landau level spacing is doubled due to the doubling of the effective electric field. } \label{fig_B1}
\end{figure*}

We now further test the chiral nature of the strain-induced pseudo-electric field $\bm e^\eta = \eta \bm e$. When an inhomogeneous electric field $\bm E$ is applied in addition to the twist, the effective electric field for the right (left) magnon Weyl cone is $\bm E_{R} = \bm E + \bm e$ ($\bm E_{L} = \bm E - \bm e$). For the special case that $\bm E = \bm e$, the left magnon Weyl cone feels no electric field but the electric field at the right magnon Weyl cone is doubled. Therefore, the left Weyl cone is unchanged but the right Weyl cone is Landau-quantized as illustrated in Fig.~\ref{fig_B1}(c). Compared to Fig.~\ref{fig_4}(c) and Fig.~\ref{fig_6}(c), the number of magnon Dirac-Landau levels in Fig.~\ref{fig_B1}(c) is halved due to the doubling of the effective electric field. These Dirac-Landau levels correspond to the even order Landau levels ($n=2, 4, \ldots$) in Fig.~\ref{fig_4}(c) and Fig.~\ref{fig_6}(c).

\section{Magnon bands of multilayer Weyl ferromagnet}
\label{a3}
In the main text, we have neglected the $J_+(1-\cos k_za) S \sigma^0$ term in the first-quantized Bloch Hamiltonian $\pazocal{H}_{\bm k}$ (Eq.~\ref{Bloch_Hk}) for ease of presentation. (We have carefully ensured that the magnon energy remains positive-definite relative to the ferromagnetic ground state.) In general, however, $J_+>0$ because the couplings $J_A$ and $J_B$ are both ferromagnetic. For this reason, we add the $J_+$ term back in this section and discuss its effects on the magnon band structure, pumping, and transport.

\begin{figure*}[htb]
\includegraphics[width = 16cm]{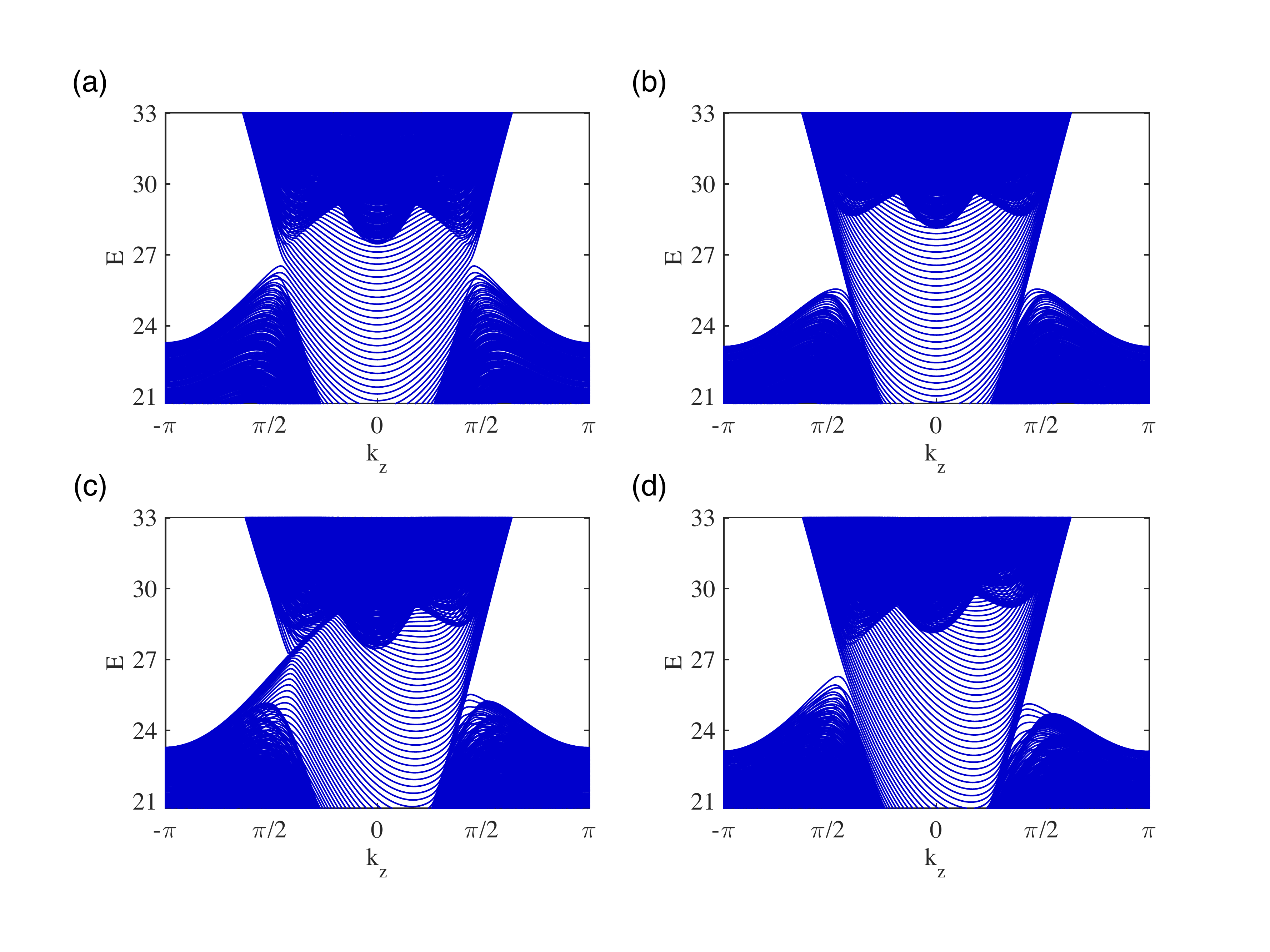}
\caption{Magnon dispersion for the Weyl ferromagnet nanowire. For all panels, the parameters are same as those of Fig.~\ref{fig_2} in the main text except that we reintroduce a nonzero $J_+ S=4.08$. (a) Magnon band structure of a nanowire without external fields. Due to the nonzero $J_+$ the Weyl cones and arc states are tilted. (b) Magnon band structure of a nanowire under an inhomogeneous external electric field whose gradient $\pazocal{E}$ satisfies $\frac{g\mu_B \pazocal{E} a^2}{ec^2} =-0.0124\Phi_0$. The Dirac-Landau levels are tilted by $J_+$ such that the velocity of the right (left) zeroth Landau level is $|v_z^\eta|+|v_0^\eta|$ ($-|v_z^\eta|-|v_0^\eta|$). (c) Magnon band structure of a twisted nanowire. The gradient of the strain-induced pseudo-electric field $\varepsilon$ satisfies $\frac{g\mu_B \varepsilon a^2}{ec^2} =-0.0124\Phi_0$. The Dirac-Landau levels are tilted by $J_+$ such that the velocity of the right (left) zeroth Landau level is $|v_z^\eta|+|v_0^\eta|$ ($|v_z^\eta|-|v_0^\eta|$). (d) Magnon band structure of a twisted nanowire under an inhomogeneous external electric field, with $\frac{g\mu_B \pazocal{E} a^2}{ec^2} = \frac{g\mu_B \varepsilon a^2}{ec^2} =-0.0124\Phi_0$. Due to the chiral nature of the strain-induced pseudo-electric field, the effective electric field at the left Weyl cone vanishes while the effective electric field at the right Weyl cone is doubled. Therefore, the left tilted Weyl cone is not Landau-quantized but the right tilted Weyl cone exhibits tilted Landau levels.} \label{fig_C1}
\end{figure*}

First, we examine the magnon band structure with the advent of the $J_+$ term. In the absence of an inhomogeneous electric field and strain, this term only shifts the magnon bands in Fig.~\ref{fig_2} by $J_+(1-\cos k_za) S$ without altering the band topology. Thus the Chern number (Eq.~\ref{Chern_number}) is still valid and guarantees that there are surface states akin to Fermi arcs connecting the magnon Weyl cones. When a transverse inhomogeneous electric field $\bm E = (\frac{1}{2}\pazocal E x, \frac{1}{2} \pazocal E y, 0)$ is switched on, Aharonov-Casher phases must be added to the magnon ``hopping" terms (the first three terms of Eq.~\ref{Bloch_Hk}). However, the diagonal term $m_{\bm k}\sigma^0 = [K_+ + J_+(1-\cos k_za) +3J_1 +6J_2]S \sigma^0$ corresponds to an ``on-site'' magnon energy whose Aharonov-Casher phase is zero; thus the $J_+$ term is invariant under the electric field. On the other hand, when a static twist is applied, an extra term $\delta \pazocal H_{\bm k}^{\bm e}$ must be added to the first-quantized Bloch Hamiltonian $\pazocal H_{\bm k}$, but as shown in Eq.~\ref{twist_dH}, $J_+$ does not contribute to $\delta \pazocal H_{\bm k}^{\bm e}$. Therefore, even in the presence of (pseudo-)electric fields, the effect of the $J_+$ term is simply shifting the magnon bands in Fig.~\ref{fig_4}, Fig.~\ref{fig_6} and Fig.~\ref{fig_B1} by $J_+(1-\cos k_za) S$. The magnon band structure in the presence of the $J_+$ term is summarized in Fig.~\ref{fig_C1}.

\begin{figure*}[htb]
\includegraphics[width = 16cm]{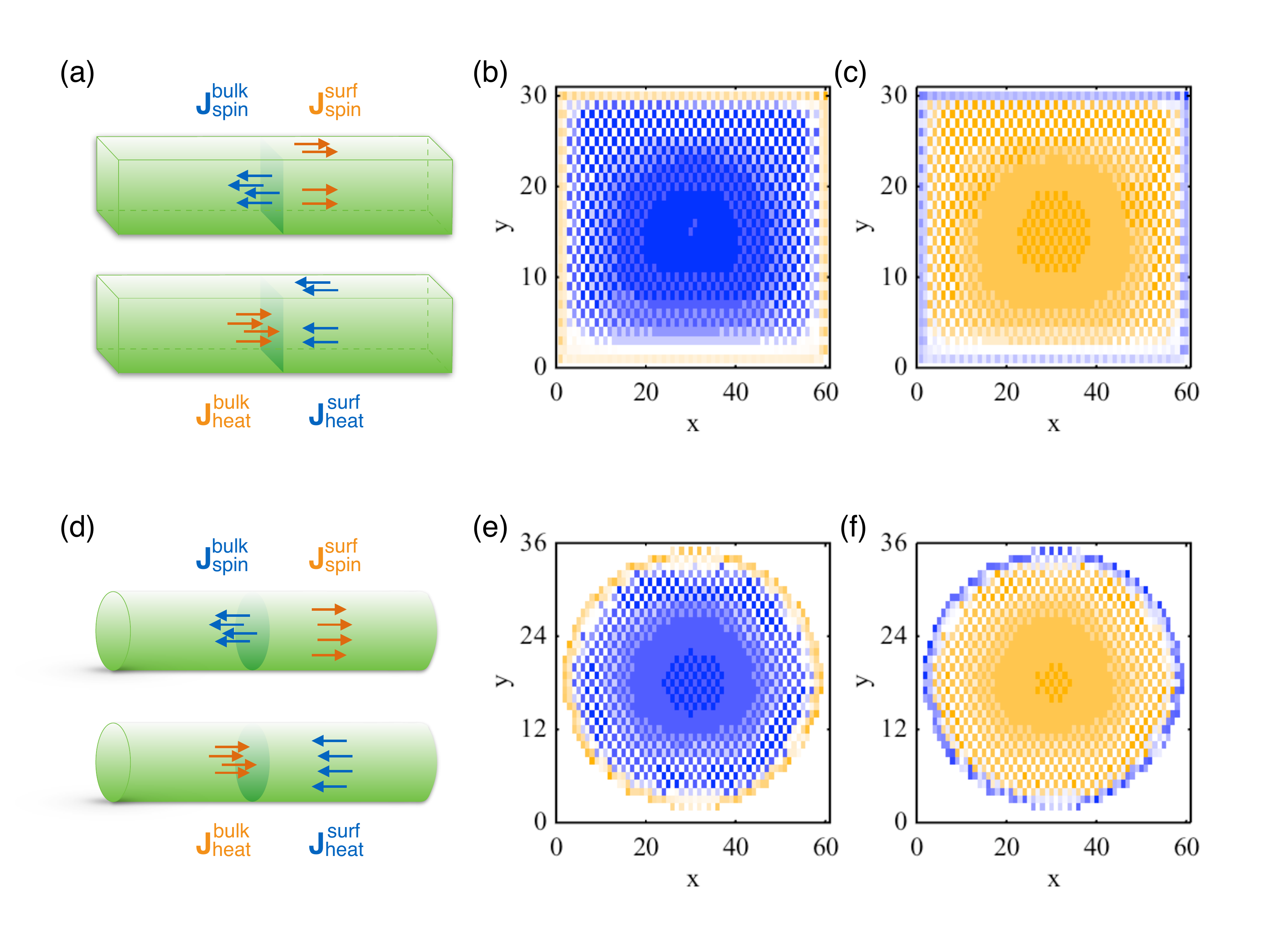}
\caption{Reproduction of bulk-surface separation for the twisted Weyl ferromagnet nanowire (Fig.~\ref{fig_8}) with a nonzero $J_+ S=4.08$. Though the Weyl cones are displaced and tilted, the bulk-surface separation of spin and heat currents is preserved for both the rectangular cross section (b, c) and the circular cross section (e, f). The total spin current on the rectangular (circular) cross section is $-0.0017DS$ ($-0.0016DS$) while the total heat current on the rectangular (circular) cross section is $0.046D^2 S^2/\hbar$ ($0.0423D^2 S^2/\hbar$).} \label{fig_C2}
\end{figure*}

Then, we discuss the effect of the $J_+$ term on the magnon equations of motion (Eqs.~\ref{EOM_BE}, \ref{EOM_bB}). In the presence of a magnetic field $\bm B$, a Zeeman energy $U=-\bm \mu \cdot \bm B$ exists for both sublattices. It is thus diagonal in the sublattice basis $\phi_{\bm k} = (a_{\bm k}, b_{\bm k})^T$, in which the Bloch Hamiltonian $\pazocal H_{\bm k}$ is defined. For this reason, the total potential energy of magnons is $U' = U +m_{\bm k}$. When $m_{\bm k}$ has no spatial dependence, the gradient of the potential energy is unchanged by $m_{\bm k}$: $\nabla U' = \nabla U$ irrespective of the value of $J_+$. Therefore, the magnon equation of motion Eq.~\ref{EOM_BE} is not affected when considering $J_+$. On the other hand, when a dynamic uniaxial strain is applied, an extra term $\delta \pazocal H_{\bm k}^{\bm b}$ must be added to the first-quantized Bloch Hamiltonian $\pazocal H_{\bm k}$. It is worth noting that the contribution from $J_+$ to $\delta \pazocal H_{\bm k}^{\bm b}$ has already been considered in Eq.~\ref{uniaxial_dH}. Therefore, the $J_+$ term is still a constant correction to the potential energy and will not affect the magnon equation of motion Eq.~\ref{EOM_bB}.

Lastly, we consider how the magnon quantum anomalies and the anomalous spin and heat currents are affected by the $J_+$ term. It is straightforward to see such a diagonal term has two effects. First, it shifts the magnon Weyl points in the energy dimension by an amount of $J_+(1-\cos Qa) S$. Second, it alters the velocities of the two zeroth Landau levels by $v_0^\eta = \eta J_+ Sa \sin Qa /\hbar$. To derive a generic theory of magnon transport in the presence of these two effects, we consider a Weyl ferromagnet nanowire aligned in the $z$ direction and subjected to a generalized electric field gradient $\mathcal{E}$, which can be generated by either an inhomogeneous electric field or a twist. Following the set-up in Sec.~\ref{s4}, the Weyl ferromagnet nanowire is attached to magnon reservoirs in a uniform magnetic field $\tilde{\bm B_0} = \tilde B_0 \hat z$ where $\tilde B_0 = B_0+J_+(1-\cos Qa) S/g\mu_B$, so that the magnon population edges on both zeroth Landau levels remains tuned into the gap spanned by the first Landau levels. Then a generalized magnetic field gradient $\mathcal{B}$ generated by either an inhomogeneous magnetic field or a dynamic uniaxial strain is applied to the nanowire, under which magnons are pumped along Landau levels according to the generalized semiclassical equation of motion
\begin{equation} \label{EOM_gen}
q_z^{R/L}(t) = q_z^{R/L}(0) - s_{R/L} g\mu_B \int_0^t \mathcal{B} dt' /\hbar,
\end{equation}
where the index $s_{R/L}$ indicates the nature of the magnetic field such that
\begin{equation}
(s_R, s_L) = 
\begin{cases}
(+1, +1) & \mbox{for magnetic field}
\\
(+1, -1) & \mbox{for pseudo-magnetic field}
\end{cases}.
\end{equation}
During the pumping process, the magnon population edge of the right/left zeroth Landau level is slightly shifted by $\delta_{R/L} \ll \tilde B_0$, so that the right/left Weyl cone experiences a magnetic field $B_{R/L} = \tilde B_0 - \delta_{R/L}$. From the generalized semiclassical equation of motion Eq.~\ref{EOM_gen}, we obtain the magnetic field variation for the right/left Weyl cone as
\begin{equation} \label{delta_RL}
\delta_{R/L} = s_{R/L}v_{R/L}\int_0^t \mathcal{B} dt'.
\end{equation}
By comparing to Eqs.~\ref{nR_EB}, \ref{nL_EB}, the magnon concentration variation on the right/left zeroth Landau level reads
\begin{multline} 
n_{R/L} = \int_{g\mu_B B_{R/L}}^{g\mu_B \tilde B_0} g_{R/L}(\epsilon) n_B(\epsilon) d\epsilon 
\\
\approx \frac{n_B(g\mu_B \tilde B_0)}{4\pi^2l_{\mathcal E}^2} \frac{g\mu_B \delta_{R/L}}{\hbar|v_{R/L}|}- \frac{n_B'(g\mu_B \tilde B_0)}{2} \frac{(g\mu_B\delta_{R/L})^2}{4\pi^2l_{\mathcal E}^2\hbar|v_{R/L}|}, 
\end{multline}
where we take the limit $g\mu_B\delta_{R/L} \ll k_BT$. The magnon density of states on the right/left zeroth Landau level is $g_{R/L}(\epsilon) = \frac{1}{2\pi l_{\mathcal E}^2} \frac{1}{2\pi \hbar |v_{R/L}|}$ with the generalized electric length $l_{\mathcal E} = (|\hbar c^2/g\mu_B \mathcal E|)^{1/2}$. Due to the variation of magnon population, thermal energy will be injected into or depleted from the right/left zeroth Landau level. Explicitly, the thermal energy variation can be obtained by referring to Eqs.~\ref{UR_EB}, \ref{UL_EB} as
\begin{multline}
U_{R/L} = \int_{g\mu_B B_{R/L}}^{g\mu_B \tilde B_0} \epsilon g_{R/L}(\epsilon) n_B(\epsilon) d\epsilon
\\ 
\approx \frac{1}{4\pi^2l_{\mathcal E}^2 \hbar |v_{R/L}|} \bigg\{ n_B(g\mu_B \tilde B_0) g\mu_B \tilde B_0 g\mu_B \delta_{R/L} \\- \frac{1}{2} [n_B(g\mu_B \tilde B_0) + g\mu_B \tilde B_0 n_B'(g\mu_B \tilde B_0)] (g\mu_B \delta_{R/L})^2 \bigg\},
\end{multline}
where we again approximate $g\mu_B\delta_{R/L} \ll k_BT$. The magnon chiral anomaly and heat anomaly to the linear order in $g\mu_B\delta_{R/L}$ are given by
\begin{multline} \label{QA_chiral}
\frac{d \rho_5}{dt} = \chi_R \frac{dn_R}{dt} + \chi_L \frac{dn_L}{dt}
\\
\approx \frac{n_B(g\mu_B \tilde B_0)}{4\pi^2l_{\mathcal E}^2 \hbar} g \mu_B \mathcal B (s_R \sgn(v_R)-s_L \sgn(v_L)),
\end{multline}
\begin{multline} \label{QA_heat}
\frac{d \rho_{\text{heat}}}{dt} =  \frac{dU_R}{dt} +  \frac{dU_L}{dt}
\\
\approx \frac{n_B(g\mu_B \tilde B_0)}{4\pi^2l_{\mathcal E}^2 \hbar} g^2\mu_B^2 \tilde B_0 \mathcal B (s_R \sgn(v_R)+s_L \sgn(v_L)).
\end{multline}
These anomalies only depend on the signs of velocities of the zeroth Landau levels. In the presence of an electric field, the inclusion of $J_+$ only changes the magnitude of $v_{R/L}$ without altering the directions of magnon propagation. Consequently, Eq.~\ref{QA_chiral} is reduced to Eq.~\ref{QA_chiral_EB} and Eq.~\ref{QA_heat} is reduced to Eq.~\ref{QA_heat_Eb}, and we still have the chiral anomaly for $(\bm E, \bm B)$ and the heat anomaly for $(\bm E, \bm b)$. In the presence of a pseudo-electric field, $J_+$ changes the velocity of the right (left) zeroth Landau level to $v_R = |v_z^\eta| + |v_0^\eta|$ ($v_L = |v_z^\eta| - |v_0^\eta|$). For a type-I Weyl ferromagnet, $|v_z^\eta| > |v_0^\eta|$, thus the advent of the $J_+$ term does not flip the sign of $v_L$. Consequently, Eq.~\ref{QA_chiral} is reduced to Eq.~\ref{QA_chiral_eb} and Eq.~\ref{QA_heat} is reduced to Eq.~\ref{QA_heat_eB}, preserving the chiral anomaly for $(\bm e, \bm b)$ and the heat anomaly for $(\bm e, \bm B)$. On the other hand, for a type-II Weyl ferromagnet, $|v_z^\eta| < |v_0^\eta|$, and the sign of $v_L$ is flipped. In this case, $(\bm e, \bm b)$ will have a heat anomaly but $(\bm e, \bm B)$ will have a chiral anomaly. The anomalous spin and heat currents can be derived as 
\begin{multline} \label{spin}
J_{\text{spin}}=-\hbar(n_Rv_R+n_Lv_L)
\\
\approx -\frac{n_B(g\mu_B \tilde B_0)}{4\pi^2l_{\mathcal E}^2 } g\mu_B [\delta_R\sgn(v_R)+\delta_L\sgn(v_L)] \\+ \frac{n_B'(g\mu_B \tilde B_0)}{8\pi^2l_{\mathcal E}^2 } g^2\mu_B^2 [\delta_R^2\sgn(v_R)+\delta_L^2\sgn(v_L)],
\end{multline}
\begin{multline} \label{heat}
J_{\text{heat}} = v_R U_R + v_L U_L
\\
\approx \frac{n_B(g\mu_B \tilde B_0)}{4\pi^2l_{\mathcal E}^2 \hbar} g^2\mu_B^2\tilde B_0 [\delta_R\sgn(v_R)+\delta_L\sgn(v_L)] 
\\ 
- \frac{n_B'(g\mu_B \tilde B_0)}{8\pi^2l_{\mathcal E}^2 \hbar} g^3\mu_B^3 \tilde B_0[\delta_R^2\sgn(v_R)+\delta_L^2\sgn(v_L)]
\\ 
- \frac{n_B(g\mu_B \tilde B_0)}{8\pi^2l_{\mathcal E}^2 \hbar} g^2\mu_B^2 [\delta_R^2\sgn(v_R)+\delta_L^2\sgn(v_L)].
\end{multline}
Unlike the anomalies (Eqs.~\ref{QA_chiral}, \ref{QA_heat}), the anomalous spin and heat currents depend on the magnitude of $v_{R/L}$ as well because $\delta_{R/L}$ is proportional to $v_{R/L}$ according to Eq.~\ref{delta_RL}. For $(\bm E, \bm B)$, both zeroth Landau levels are steeper when the diagonal term is considered ($v_R = -v_L = |v_z^\eta|+|v_0^\eta|$). Therefore, the anomalous spin and heat currents are enhanced and the chiral electric effect becomes more pronounced. For $(\bm E, \bm b)$, both the spin current and the heat current are zero because there are an equal number of right-moving magnons at the speed of $v_R= |v_z^\eta|+|v_0^\eta|$ and left-moving magnons at the speed of $v_L= -|v_z^\eta|-|v_0^\eta|$. For $(\bm e, \bm b)$, importantly, the terms linear in $\delta_{R/L}$ in Eqs.~\ref{spin}, \ref{heat} are non-vanishing regardless of the sign of $v_{R/L}$, because $|\delta_{R}| \neq |\delta_L|$ when the $J_+$ term is considered. Therefore, the anomalous spin/heat current $J_{\text{spin/heat}}^{\bm e, \bm b} \sim b_5^2$ (Eqs.~\ref{spin_eb}, \ref{heat_eb}) will be obscured by the dominant linear terms. Nevertheless, these less dominant chiral electric currents can in principle be extracted because they have unique pseudo-EM field dependence. For $(\bm e, \bm B)$, the chiral torsional spin/heat current $J_{\text{spin/heat}}^{\bm e, \bm B} \sim \delta_B$ (Eqs.~\ref{spin_eB}, \ref{heat_eB}) is linear in the magnon population edge variation $\delta_B$, which is now replaced by $\frac{1}{2}[\delta_R\sgn(v_R)+\delta_L\sgn(v_L)]$ in Eqs.~\ref{spin}, \ref{heat}. Such a variation, however, effects no qualitative changes to anomalous currents. To prove this, we reproduce Fig.~\ref{fig_8} with the $J_+$ term considered. As illustrated in Fig.~\ref{fig_C2}, for a Weyl ferromagnet nanowire with a either rectangular or circular cross section, the bulk-surface separation for anomalous currents persists.

\section{Circular bend induced pseudo-electric field}
\label{a4}
In Sec.~\ref{s6} of the main text, we propose that the magnon chiral anomaly (Eq.~\ref{QA_eb}) and the magnon heat anomaly (Eq.~\ref{QA_Eb}) may have experimentally measurable mechanical effects if an additional chiral pseudo-electric field $\bm e'_\eta = \eta (0, \varepsilon' z, 0)$ is applied. In this section, we will elaborate the implementation of such an additional pseudo-electric field.

\begin{figure}[htb]
\includegraphics[width = 7cm]{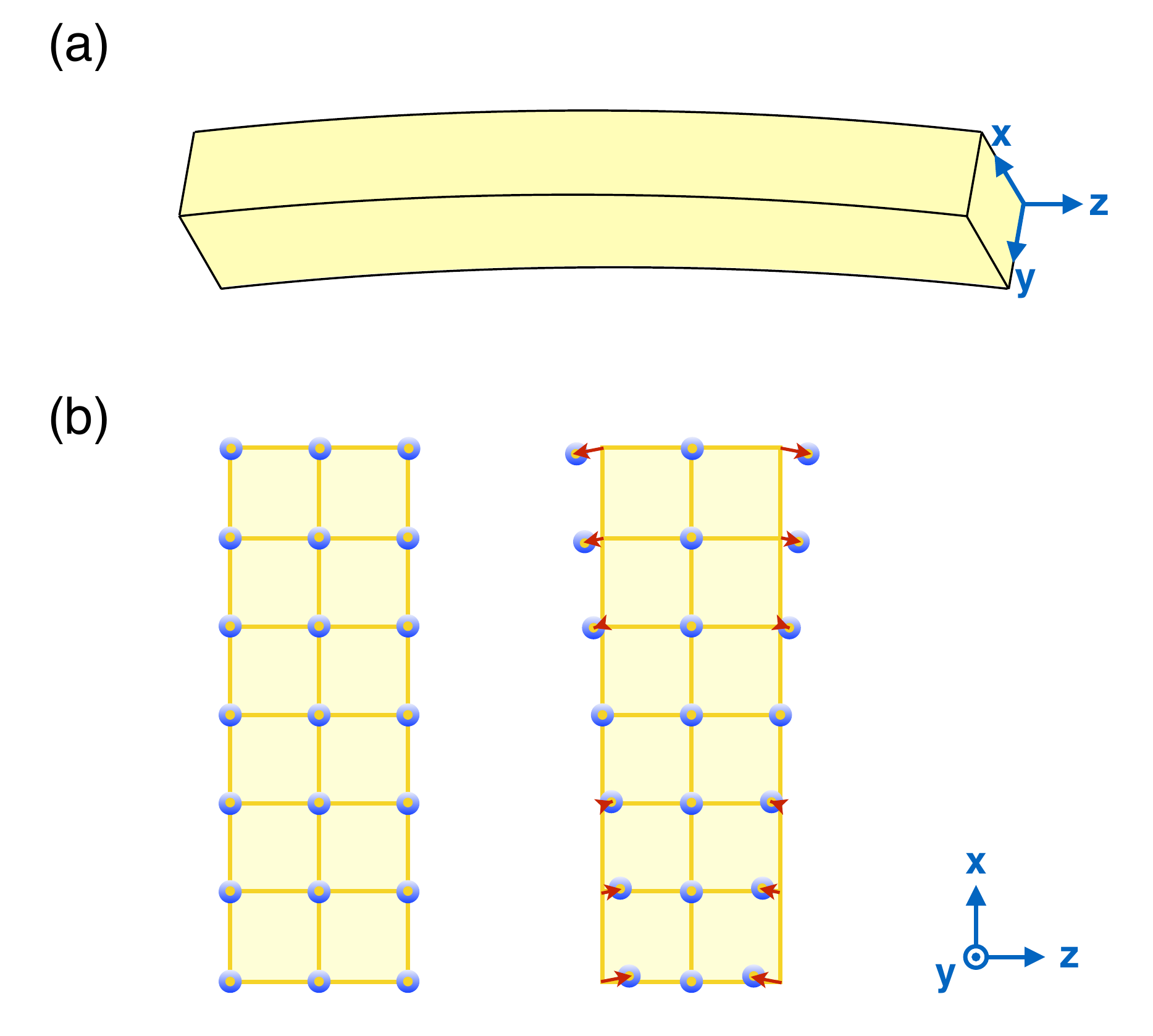}
\caption{Schematic plot for the Weyl ferromagnet nanowire. (a) Nanowire under a circular bend deformation. (b) Lattice site positions without deformation (left) and with a circular bend (right). } \label{fig_D1}
\end{figure}

We consider a simple circular bend lattice deformation as illustrated in Fig.~\ref{fig_D1}. As explained in Refs.~\cite{liu2017a, liu2017b}, to the lowest order of approximation, the displacement field of such a circular bend is $\bm u = \mathcal Uxz \hat z$, resulting in nonzero strain tensor components $u_{13} = u_{31} = \frac{1}{2} \mathcal Uz$ and $u_{33} = \mathcal Ux$. According to Eq.~\ref{sub_gen}, the strain effect can be incorporated by the following exchange integral substitutions
\begin{align*}
J_2(\bm \alpha_1 \pm a \hat z) & \rightarrow J_2 ( 1\mp\tfrac{\sqrt{3}}{2}u_{31} - \tfrac{1}{2} u_{33} ),
\\
J_2(\bm \alpha_2 \pm a \hat z) & \rightarrow J_2 ( 1 \pm\tfrac{\sqrt{3}}{2}u_{31} - \tfrac{1}{2} u_{33} ),
\\
J_2(\bm \alpha_3 \pm a\hat z) & \rightarrow J_2 (1 - \tfrac{1}{2} u_{33}),
\\
J_A & \rightarrow J_A (1-u_{33}),
\\
J_B & \rightarrow J_B (1-u_{33}),
\end{align*}
which result in an effective Hamiltonian
\begin{multline}
\pazocal H_{\bm k_W^\eta + \bm q} + \delta \pazocal H_{\bm k_W^\eta + \bm q} \approx \pazocal H_{\bm k_W^\eta} + \hbar v_0^\eta \Big( q_z + \frac{e}{\hbar}  a_z^\eta \Big) \sigma^0
\\
+ \sum_i \hbar v_i^\eta \Big( q_i + \frac{e}{\hbar} a_i^\eta \Big) \sigma^i - 3 J_2 S u_{33} \sigma^0.
\end{multline}
Here the strain-induced vector potential is
\begin{equation}
\bm a^\eta = -\eta \frac{\hbar}{ea} \bigg(\frac{2J_2 \sin Qa}{J_1 + 2J_2 \cos Qa} u_{31}, 0, \frac{1-\cos Qa}{\sin Qa} u_{33} \bigg),
\end{equation}
which is incorporated by a minimal substitution in the same way as $\bm a_S^\eta$ and $\bm a_D^\eta$. Again, a non-chiral on-site term $-3J_2Su_{33}$ appears, but such a term is negligible when $\hbar |v_z^\eta|/a \gg J_2S$, and it can be otherwise cancelled by an additional magnetic field $\bm B' = -\frac{3J_2Su_{33}}{\mu^2}\bm \mu$. Therefore, its effect on magnon mechanics can be safely neglected. On the other hand, by referring to Eq.~\ref{EOM_bB}, the chiral gauge vector potential $\bm a^\eta$ gives rise to a magnon Lorentz force
\begin{equation}
\hbar \frac{d \bm k}{dt} = -e \bm v \times ( \nabla \times \bm a^\eta). 
\end{equation}
Comparing to the magnon Lorentz force (Eq.~\ref{EOM_Ep}) due to an additional electric field $\bm E' = (0, \pazocal E' z, 0)$, we may interpret $\bm a^\eta$ as an additional chiral pseudo-electric field $\bm e'_\eta = \eta(0, \varepsilon' z, 0)$, which only differs from $\bm E'$ by a chiral charge $\eta$. The gradient of this additional pseudo-electric field can be determined by $\nabla \times \bm a^\eta = \frac{1}{ec^2} \nabla \times (\bm e'_\eta \times \bm \mu) = - \eta \frac{g\mu_B \varepsilon'}{ec^2} \hat y$. Explicitly, the field gradient reads
\begin{multline}
\varepsilon' = -\eta \frac{ec^2}{g\mu_B} \hat y \cdot (\nabla \times \bm a^\eta) 
\\
= \frac{\hbar c^2}{g\mu_Ba} \bigg( \frac{J_2 \sin Qa}{J_1+2J_2\cos Qa} - \frac{1-\cos Qa}{\sin Qa} \bigg) \mathcal U.
\end{multline} 
As discussed in Sec.~\ref{s6b} (Sec.~\ref{s6c}), when $\varepsilon'$ is sufficiently small, i.e., $\varepsilon' \ll \varepsilon$ ($\varepsilon' \ll \pazocal E$),  the bend-induced pseudo-electric field $\bm e'_\eta$ generates a transverse force that may be measured by AFM, while the magnon Landau levels  are not strongly affected. On the other hand, if $\varepsilon'$ is sufficiently large, the bend-induced pseudo-electric field $\bm e'_\eta$ also results in Landau quantization. In contrast to the Landau quantization due to a twist, the Landau quantization due to a bend takes place in the transverse direction.

\bibliography{a1}

\begin{thebibliography}{63}
\bibitem[{not({\natexlab{a}})}]{note4}
\bibinfo{note}{At least in principle, there is another proposal to
  experimentally realize such electric field. In particular, a uniformly
  charged cylindrical Weyl ferromagnet nanowire produces an electric field $\bm
  E = \frac{1}{2} \pazocal E r \hat r$, where the field gradient is $\pazocal E
  = \frac{\rho}{\epsilon_0 \epsilon_r}$ with $\rho$ being the charge density
  and $\epsilon_0\epsilon_r$ being the permittivity of the nanowire. For the
  field gradient used in Fig.~\ref{fig_4}, a charge density $\sim
  10^{10}$C$/$m$^3$ is needed.}
\bibitem[{not({\natexlab{b}})}]{note3}
\bibinfo{note}{In systems where the non-chiral gradient term is no longer
  negligible, its effects are similar to those of an inhomogeneous magnetic
  field $\bm B$. Therefore, the resulting magnon quantum anomalies will be a
  combination of purely magnetic field induced anomalies and purely chiral
  pseudo-magnetic field induced anomalies.}
\bibitem[{not({\natexlab{c}})}]{note1}
\bibinfo{note}{The minus sign comes from the assumption that
  $\sgn(g\pazocal{E})=-1$ in Sec.~\ref{s3a}.}
\bibitem[{not({\natexlab{d}})}]{note2}
\bibinfo{note}{In the limit $g\mu_B\delta_b \ll k_BT$, the total magnon
  concentration variation on the zeroth Landau levels $n_R^{\bm E, \bm
  b}+n_L^{\bm E, \bm b} \sim g\mu_B\delta_b$ is linear in the magnetic field
  variation $\delta_b$. This is fundamentally different from the total magnon
  concentration variation $n_R^{\bm E, \bm B}+ n_L^{\bm E, \bm B} \sim
  g^2\mu_B^2B_5^2$ ($n_R^{\bm e, \bm b}+ n_L^{\bm e, \bm b} \sim
  g^2\mu_B^2b_5^2$) associated with magnon chiral anomaly in the low bias
  limit, which is quadratic in the magnetic field bias $B_5$ ($b_5$).}
\end{thebibliography}
\end{document}